\documentclass[fleqn,useAMS]{mnras}
\usepackage{newtxtext,newtxmath}

\usepackage{graphicx}	
\usepackage{amsmath}	
\usepackage{multicol}        
\usepackage{bm}		
\usepackage{pdflscape}	

\usepackage[normalem]{ulem} 
\usepackage[dvipsnames]{xcolor}
\usepackage{soul}

\usepackage{tikz}
\usetikzlibrary{calc, positioning}

\usepackage[percent]{overpic}

\usepackage{booktabs, multirow, makecell}

\setcellgapes{5pt}
\newcommand{\rh}{\, r_{\rm half}}
\newcommand{\msol}{\, M_{\odot}}

\newcommand{\kms}{\,{\rm km \, s^{-1}}}
\newcommand{\pc}{\,{\rm pc}}
\newcommand{\kpc}{\,{\rm kpc}}

\newcommand{\oversim}[2]{\protect{\mbox{\lower0.5ex\vbox{%
   \baselineskip=0pt\lineskip=0.2ex
   \ialign{$\mathsurround=0pt #1\hfil##\hfil$\crcr#2\crcr\sim\crcr}}}}} 
\newcommand{\bb}[1]{\ifmmode \mbox{\boldmath $ #1$} \else  \mbox{\boldmath $#1$} \fi}
\catcode`\"=\active\let"=\"  

\def\3{{\ss} }

\def\c12{{1\over 2}}

\def\d{{\rm d}}   
   
\def\plusplus{\raise 0.3ex\hbox{${\scriptstyle ++}$}{}}

\def\and{{{\rm M}31}}

\def\gyr{\,{\rm Gyr}}

\usepackage{array}
\newcolumntype{L}[1]{>{\raggedright\let\newline\\\arraybackslash\hspace{0pt}}m{#1}}
\newcolumntype{C}[1]{>{\centering\let\newline\\\arraybackslash\hspace{0pt}}m{#1}}
\newcolumntype{R}[1]{>{\raggedleft\let\newline\\\arraybackslash\hspace{0pt}}m{#1}}

\usepackage[normalem]{ulem}

\pdfoutput=1

\DeclareGraphicsExtensions{.pdf,.png,.jpg}
\begin{document}
\title[Gravothermal expansion of dSphs]{Gravothermal expansion of dwarf spheroidal galaxies heated by dark subhaloes}
\author[Pe\~narrubia et al.]{Jorge Pe\~narrubia$^{1,2}$\thanks{Email: jorpega@roe.ac.uk}, Rapha\"el Errani$^{3}$, Eduardo Vitral$^{1}$, Matthew G. Walker$^{3}$\\
$^1$Institute for Astronomy, University of Edinburgh, Royal Observatory, Blackford Hill, Edinburgh EH9 3HJ, UK\\
$^2$Centre for Statistics, University of Edinburgh, School of Mathematics, Edinburgh EH9 3FD, UK\\
$^3$McWilliams Center for Cosmology and Astrophysics, Department of Physics, Carnegie Mellon University, Pittsburgh, PA 15213, USA\\}
\maketitle
\begin{abstract}
  We use analytical and $N$-body methods to study the evolution of dwarf spheroidal galaxies (dSphs) embedded in dark matter (DM) haloes that host a sizeable subhalo population. Dark subhaloes generate a fluctuating gravitational field that injects energy into stellar orbits, driving a gradual expansion of dSphs.
  Despite the overall expansion, the stellar density profile preserves its initial shape, suggesting that the evolution proceeds in a self-similar manner. Meanwhile, the velocity dispersion profile, initially flat, evolves as the galaxy expands: the inner regions heat up, while the outer regions cool down. Kinematically, this resembles gravothermal collapse but with an inverted evolution, instead of collapsing the stellar system expands within a fluctuating halo potential.
As the half-light radius $\rh$ approaches the halo peak velocity radius $r_{\rm max}$, the expansion slows, and the velocity dispersion peaks at $\sigma_{\rm max} \simeq 0.54 v_{\rm max}$. The stellar heat capacity remains positive for deeply embedded stars but diverges near $r_{\rm max}$, turning negative beyond this threshold, which indicates a phase transition in the dynamical response to energy injection.  
  The relaxation time scales as $t_{\rm rel} \sim \rh^{3/2}$, showing that orbital diffusion slows as the galaxy expands. Ultra-faint dSphs, having the smallest sizes and shortest relaxation times, are particularly sensitive to the presence of dark subhaloes. 
  Some of our dSph models expand beyond the detection of current photometric surveys, becoming `stealth' galaxies with luminosities and metallicities akin to known ultrafaints but with larger sizes and higher velocity dispersions. These objects would display half-light radii and dispersions similar to ultra-diffuse galaxies, but remain orders of magnitude fainter, representing a distinct, yet currently undetected, population of DM-dominated satellites.

\end{abstract}


\section{Introduction}\label{sec:intro}
According to the standard cold dark matter (CDM) model, all galaxies are embedded in dark matter haloes that contain a large population of subhaloes (e.g. Klypin et al. 1999; Moore et al. 1999). Most of these subhaloes remain `dark' throughout their evolution, forming no stars (e.g. Benitez-Llambay \& Frenk 2020). Detecting or ruling out their existence is a key goal in modern cosmology, as their abundance and properties provide crucial insights into the nature of dark matter (see Bullock \& Boylan-Kolchin 2017 for a review).

Several observational techniques have been proposed to probe the subhalo population. These include the Lyman-$\alpha$ forest power spectrum (e.g. Viel et al. 2013), flux ratios in gravitationally lensed quasars (e.g. Mao \& Schneider 1998), gamma-ray sources (e.g. Zechlin et al. 2012; Amerio et al. 2025), and perturbations in strong lensing arcs (e.g. Koopmans 2005; Nadler et al. 2021; Despali et al. 2022; O'Riordan et al. 2023). Other dynamical tracers include the survival of wide-separation stellar binaries (Pe\~narrubia et al. 2010b) and heating of cold stellar streams (e.g. Ibata et al. 2002; Erkal \& Belokurov 2015; Lovell et al. 2022; Zhang et al. 2025). Indirect methods have also been proposed, such as circumgalactic gas compression near subhaloes (McCarthy \& Font 2020) or microlensing of ionized interstellar material (Delos 2024).

Dwarf spheroidal galaxies (dSphs), as the most dark matter-dominated galaxies in the universe, provide a unique laboratory for testing DM theories. Their old stellar populations and low velocity dispersions make them highly sensitive to perturbations from dark subhaloes. Recently, Pe\~narrubia et al. (2024) demonstrated that subhaloes with masses $M\gtrsim 10^5\msol$ can leave observable signatures in dSphs by capturing field stars onto temporarily bound orbits, producing spatial and kinematic overdensities. Motivated by this result, this paper explores the impact of a large population of dark subhaloes on the internal dynamics of dSphs.

There are two major sources of uncertainty in our study. 
First, the masses and scale radii of dSph host haloes are largely unknown. 
Cosmological simulations show that dSph galaxies must form in haloes with masses $M_h/\msol\sim 10^9$--$10^{10}$ in order to reproduce the observed scaling relations between size, velocity dispersion and luminosity (e.g. Pe\~narrubia et al. 2008a; Errani et al. 2018; Genina et al. 2023). However, on an individual-galaxy basis, stellar kinematics mostly constrain the amount of DM within the half-light radius (e.g. Walker et al. 2009; Wolf et al. 2010; Amorisco \& Evans 2011; Campbell et al. 2017). As a result, estimates of the total mass and scale radius for individual dwarf galaxies typically rely on uncertain extrapolations of the DM density profile to regions devoid of stars. Even if such extrapolations are accurate, the limited kinematic data and lack of full phase-space information introduce degeneracies (e.g. between enclosed mass and velocity anisotropy) that significantly hinder precise measurements of the DM distribution in dSphs.

Second, the number of subhaloes in the MW dSphs remains a fully open question. Cosmological simulations struggle to make reliable predictions in this regime due to numerical limitations. First, identifying subhaloes in dense environments is difficult, as traditional methods such as the Friends-of-Friends (FoF) algorithm (e.g. Press \& Davis 1982) often fail to distinguish satellites within hosts. The problem worsens dramatically when attempting to resolve sub(sub)haloes within subhaloes. Many algorithms have been developed with the aim of finding subhaloes that go beyond the FoF technique. However, existing methods typically return discrepant results (Forouhar-Moreno et al. 2024; Mansfield et al. 2024; Diemer et al. 2024).
In addition, the finite resolution of cosmological codes produces numerical artifacts that can lead to artificial disruption of subhaloes (van den Bosch \& Ogiya 2018; van den Bosch et al. 2018; Errani \& Pe\~narrubia 2020). Resolving tidally-processed subhaloes requires excellent numerical resolution and a large number of particles per subhalo. Poor spatial resolution (e.g., grid size or `softening-length'), and/or poor time resolution (e.g., timestep) systematically lower the characteristic subhalo densities and make them prone to artifical disruption (Errani et al. 2023). High-resolution controlled simulations indicate that cuspy subhaloes must be resolved with at least 3000 particles within the peak velocity radius to achieve numerical convergence (Errani \& Navarro 2021). On the other hand, cosmological simulations of Milky Way-like galaxies find that the number of surviving self-bound subhaloes increases with resolution (e.g. Santos-Santos et 2024). Accurately tracking the dynamical evolution of sub(subhaloes) in dSphs within a cosmological setting remains a formidable numerical challenge, and an open problem in computational cosmology. In contrast, semi-analytical models have made substantial progress in predicting substructure demographics (e.g. Jiang \& van den Bosch 2016, 2017). These studies suggest that higher-order substructure (i.e., sub-subhaloes, sub-sub-subhaloes, etc.) exhibit self-similar properties: their mass and size functions follow universal distributions when expressed in units of the mass and scale radius of their parent haloes.

 This self-similarity is used in this paper to build controlled $N-$body experiments of dSphs that host sizeable subhalo populations. In these models, subhaloes are rescaled from those found in the Aquarius simulations (Springel et al. 2008) to match expectations for the dark matter haloes of Milky Way dSphs.
 
 In \S\ref{sec:toy}, we construct toy models of the masses, sizes, and orbits of sub(sub)haloes in these galaxies. In \S\ref{sec:evol}, we study the motion of stellar tracers in DM haloes containing a large population of subhaloes. We will show that these objects generate a fluctuating force field that gradually injects energy into stellar orbits, leading to progressive expansion of the galaxy. In \S\ref{sec:sto}, we employ virial equations and Chandrasekhar's stochastic theory to model the kinematic evolution of expanding dSphs. \S\ref{sec:dis} discusses the limitations of our models and the implications of our results in an observational context. \S\ref{sec:sum} provides a summary of the main findings of the paper .

\section{A toy model for dark subhaloes in dwarf spheroidals}\label{sec:toy}

The primary goal of this paper is to analyze the evolution of dSphs that host a population of subhaloes.
Rather than constructing fully `realistic' models of Milky Way dSphs, which would require a detailed study accounting for their hierarchical assembly and evolution within the Galactic tidal field, we instead design controlled statistical experiments to address the following theoretical questions:
\begin{itemize}
    \item How do dark subhaloes influence the internal dynamics of dSphs?
    \item Does their effect depend on the size and velocity dispersion of the host dSph?
    \item What role does the subhalo mass function play?
\end{itemize}

To investigate these questions, we take the following approach. First, we model host dark matter haloes to reproduce the observed scaling relations of Milky Way dSphs. Second, we generate subhalo populations with masses and sizes consistent with those found in CDM simulations of structure formation. Finally, we assign these subhaloes random orbits within the host halo in dynamical equilibrium. Our experiments are highly idealized and rest on a number of simplifying assumptions that are outlined below for clarity.

\subsection{Model assumptions}\label{sec:assumptions}
According to the standard cosmological paradigm (CDM), dSphs are embedded in dark matter haloes that grew through the accretion of smaller subhaloes. Once accreted, these subhaloes orbit within the dSph potential, with their apocentres scaling approximately with the virial radius of the parent halo at the redshift of accretion. At fixed halo mass:
  $$  r_{\rm apo} \simeq r_{200}(z_{\rm acc}) = r_{200}(z=0)(1+z_{\rm acc})^{-1}$$
(e.g., Bullock et al. 2001; Errani et al. 2017, 2024; Santos-Santos et al. 2024). Thus, subhaloes accreted at higher redshifts tend to have smaller apocentres, and higher orbital frequencies than those accreted later\footnote{This picture is somewhat oversimplified. The orbits of subhaloes evolve with time as the host halo grows hierarchically (e.g. van den Bosch et al. 2016; Santistevan et al. 2023).}

The accretion of new subhaloes ceases soon after dSph galaxies become satellites of the Milky Way. As they orbit around the larger host, dSphs gradually lose a fraction of their subhaloes to Galactic tides, particularly after each pericentric passage. Consequently, the subhaloes that remain bound to dSphs at $z=0$ are likely those that were tightly bound at their time of accretion. Since stars occupy the inner-most regions of the host halo, most interactions between stars and dark subhaloes involve DM substructures that have been orbiting in the innermost regions of dSphs for extended periods of time.

Because the tidal field of a dSph is strongest near its centre, and higher orbital frequencies result in a greater number of orbital revolutions within a Hubble time, we expect the present population of subhaloes to have undergone significant tidal evolution. That is, they should exhibit structural properties shaped by repeated tidal mass loss and subsequent re-virialization within their host halo. As demonstrated by controlled $N$-body simulations, tidally stripped CDM subhaloes follow well-defined evolutionary tracks (e.g., Hayashi \& Navarro 2002; Pe\~narrubia et al. 2008b,2010a; Errani \& Navarro 2021; Green \& van den Bosch 2022; Du et al. 2024) and their density profiles asymptotically approach an exponentially truncated NFW cusp (Errani \& Navarro 2021).

The experiments presented below adopt an idealized set-up that seeks to reproduce the overall characteristics of subhaloes that one may expect to find in cosmological simulations of structure formation. To this aim, we introduce the following ansatzs:
\begin{itemize}
  \item Subhaloes in dSphs are `dark' (i.e. they do not form stars in-situ). 
  \item Their mass function and number density distribution are separable.
  \item They are in dynamical equilibrium within the dSph galaxy potential.
  \item Their orbital velocities  can be modelled with an Opsikov-Merritt (1985) distribution function, which is isotropic at small radii and becomes radially anisotropic at large radii.
      \item The number density of subhaloes follows the dark matter distribution in the host dSph.
  \item Their individual density profiles have asymptotically converged to exponentially-truncated NFW haloes with characteristic densities that are determined by the mean density of the host halo at their orbital pericentres.
\end{itemize}
These assumptions simplify the analysis in several ways: (i) The subhalo mass function remains spatially invariant across the host galaxy. (ii) The orbits of subhaloes do not evolve over time. (iii) Once subhaloes reach an asymptotic equilibrium, their internal structure remains fixed, even as they move on eccentric orbits.

Importantly, under these conditions, the structure of subhaloes is dictated solely by the background density profile of the host dSph along their orbits. This means that they have lost memory of their individual evolutionary histories within the host dSph.

\subsection{The DM haloes of dSphs}\label{sec:haloes}

For analytical convenience, the DM haloes of dSphs are assumed to follow Hernquist (1990) profile 
\begin{align}\label{eq:hern}
  \rho(r)=\frac{\rho_0}{(r/c_h)(1+r/c_h)^3},
\end{align}
where $\rho_0=M_h/(2\pi c_h^3)$ is a characteristic density, $M_h$ is the total mass and $c_h$ is the scale radius of the halo. As the universal profile found in CDM simulations of structure formation (e.g. Navarro, Frenk \& White 1997, hereafter NFW), Equation~(\ref{eq:hern}) exhibits a centrally-divergent density cusp, $\rho\sim r^{-1}$ at small radii, $r\ll c_h$. At large distances, $r\gg c_h$, Equation~(\ref{eq:hern}) follows as a power-law $\rho \sim r^{-4}$, which provides a better match to the steeper profiles of tidally-processed satellite haloes (e.g. Pe\~narrubia et al. 2009). 

The stellar number density profiles of dSphs are typically fitted with Plummer (1905) models
\begin{align}\label{eq:plummer}
  \nu(r)=\frac{\nu_{0}}{[1+(r/a)^2]^{5/2}},
\end{align}
where $a$ is the scale radius, and the half-light radius is $\rh=1.305\,a$. In projection, the half-light radius is equal to the Plummer scale radius, $R_{\rm half}=a$. In what follows, we assume that the contribution from the stellar mass to the dSph potential can be neglected (Simon 2019 and references therein). It is therefore convenient to normalize the distribution~(\ref{eq:plummer}) to unity by choosing $\nu_0=(4\pi \,a^3/3)^{-1} $.

In equilibrium, the luminosity-averaged 1D velocity dispersion of tracer stars embedded in an extended DM halo can be computed from the virial theorem as $\sigma^2=-W$, where 
\begin{align}\label{eq:W}
  W=-\frac{4\pi }{3}\int_0^\infty \d r r^2\,\nu(r) \frac{G M_h(<r)}{r},
\end{align}
is the potential energy of stellar tracers moving in the DM halo (e.g. Errani et al. 2018).

It is well known that the half-light radius ($\rh$) and the luminosity-averaged dispersion ($\sigma$) do not provide sufficient information to constrain the {\it individual} masses of the DM haloes that host the MW dSphs. In particular, a strong degeneracy arises between the segregation of the stars within the halo ($\rh/c_h$), and the halo mass ($M_h$), see e.g. Pe\~narrubia et al. (2008a).  On an individual galaxy basis, the aforementioned degeneracy can be broken by restricting the allowed range of masses and scale radii of the haloes of dSphs to those of the subhaloes found in cosmological simulations of Milky Way-like galaxies (e.g. Errani et al. 2018).

\begin{figure}
\begin{center}
\includegraphics[width=84mm]{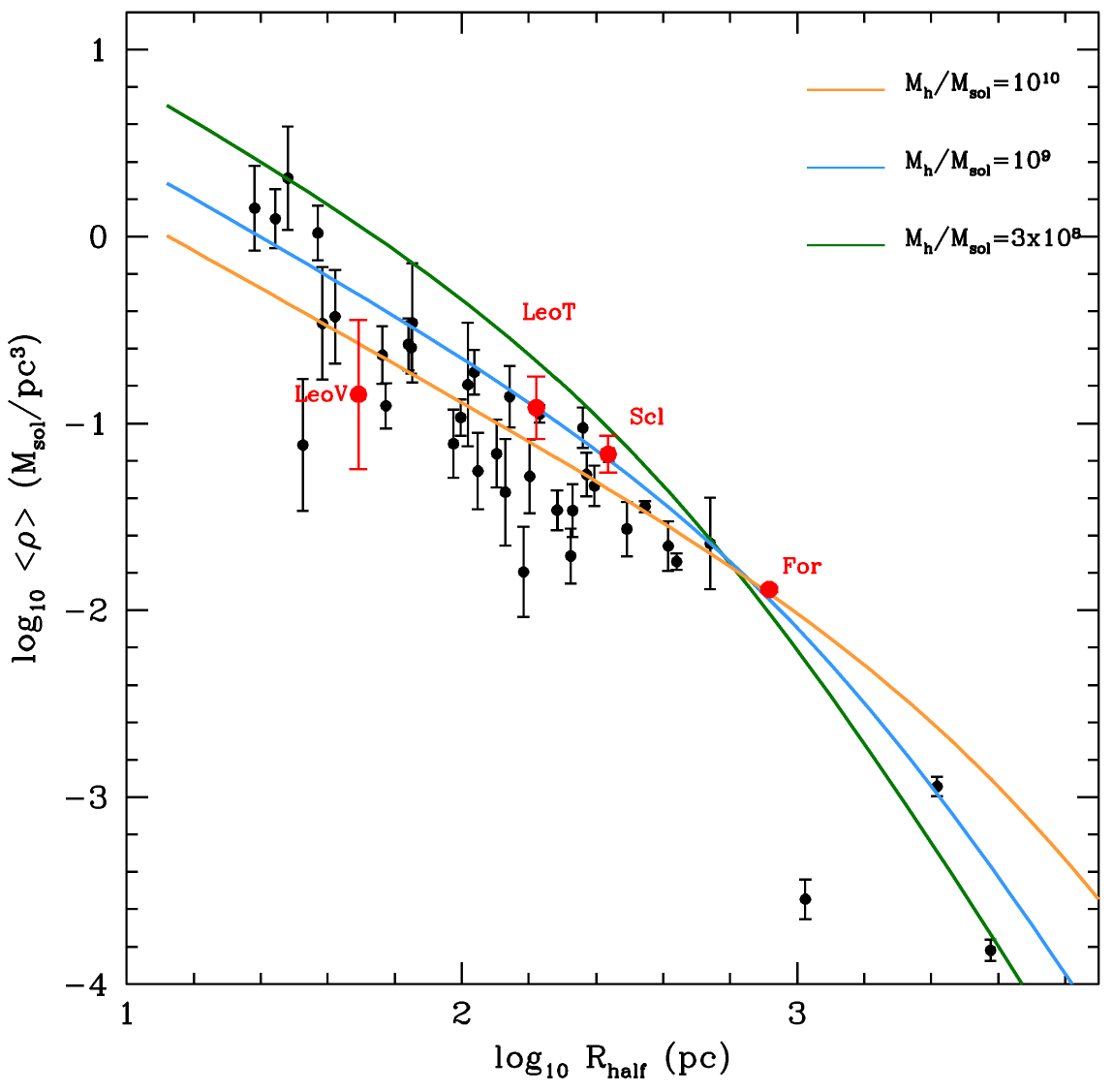}
\end{center}
\caption{Projected half-light radius ($R_{\rm half}$) versus mean densities ($\langle \rho\rangle$) of Milky Way dSphs with available velocity dispersion. To compute the mean densities, $\langle \rho\rangle = 3\,M_h(<r')/(4\pi r'^3)$, we use the enclosed mass estimated as $M_h(<r')=3.5\times r' \,\sigma^2/G$ (Errani et al. 2018), where $r'=1.8\,R_{\rm half}$ and $\sigma$ are the half-light radius luminosity-averaged velocity dispersion, respectively.
  Coloured lines show the mean densities of Hernquist haloes with masses $M_h/M_\odot=3\times 10^8,10^9$ and $10^{10}$, with scale radii $c_h/\kpc=0.75, 2.26$ and 9.95, respectively. In red we highlight 4 dSphs that will be used in the remainder of this paper for reference.}
\label{fig:rhdens}
\end{figure}

Alternatively, one may look for a unique DM halo that fits the observed relation between the half-light radius and velocity dispersion of {\it all} dSphs (e.g. Walker et al. 2009). This is illustrated in Fig.~\ref{fig:rhdens}, which shows the variation of the mean density profiles of three DM haloes with masses $M_h/M_\odot=3\times 10^8, 10^9$ and $10^{10}$ as a function of radius. The respective scale radii, $c_h/\kpc=0.75, 2.26$ and 9.95, were chosen to match the mean density of Milky Way dSphs with known velocity dispersion (black dots). Here, the mean density is derived from the virial mass estimator proposed by Errani et al. (2018), which combines the half-light radius ($\rh$) and the luminosity-averaged velocity dispersion ($\sigma$) to estimate the DM mass enclosed within a radius $r'=1.8\,R_{\rm half}$ as $M(<r')=3.5\,r'\sigma^2/G$. The mean density then follows as $\langle \rho\rangle =3M(<r')/(4\pi r'^3)$.

Fig.~\ref{fig:rhdens} reveals a clear correlation between galaxy size and DM density, where galaxies with small half-light radii are systematically denser than than those with large half-light radii\footnote{The data can be found at the Local Volume data base from Pace et al. (2024).}. 
The estimated densities agree with the picture where dSphs are currently embedded in DM haloes comprising a relatively narrow range of halo masses, $10^8\lesssim M_h/M_\odot\lesssim 10^{10}$ (e.g. Strigari et al. 2007; Pe\~narrubia et al. 2008; Kravtsov 2010; Errani et al. 2018). A noticeable exception is the Crater II dSph, an enormous dwarf galaxy with $R_{\rm half}\simeq 3800\pc$ and an extremely low density, $\langle \rho\rangle \approx 4\times 10^{-4}M_\odot/\pc^3$, which is difficult to explain by CDM models (Torrealba et al. 2016). It is also interesting to note that the densities of dSphs with sizes $R_{\rm half}\lesssim 500\pc$ are best matched by high-mass haloes, whereas large dSphs with $R_{\rm half}\gtrsim 1000\pc$ prefer low-mass halo models. Given that the luminosity of dSphs is positively correlated with their half-light radius, this trend is difficult to reproduce using simplistic galaxy formation models (see also Errani et al. 2022). 

\subsection{Subhalo populations}\label{sec:subsubs}
To generate populations of subhaloes in the dark matter (DM) haloes of dSphs, we scale the subhalo mass function from the Aquarius simulation (Springel et al. 2008) to the halo masses estimated in \S\ref{sec:haloes}. Following Han et al. (2016), we express the number density of subhaloes in the mass range $M,M+\d M$ within a volume element $r,r+\d^3r$ as
\begin{align}\label{eq:dndm}
\frac{\d n}{\d M}=\frac{\d^4 N}{\d M\d^3 r}=\frac{\d N}{\d M}g(r),
\end{align}
where
\begin{align}\label{eq:dNdM}
\frac{\d N}{\d M}=A_0\bigg(\frac{M}{M_0}\bigg)^{-\alpha},
\end{align}
is a power-law mass function with index $\alpha=1.9$ (Springel et al. 2008), $A_0$ is an arbitrary normalization, $M_0$ is the mass unit, and $g(r)$ describes the spatial number density profile of subhaloes, which is normalized to unity for convenience. The subhalo mass function is defined within a mass range $(M_1,M_2)$. Physically, the low-mass end of the mass function ($M_1$) is mainly set by the DM particle mass (e.g. Green et al. 2005), and is therefore a parameter of interest in our study.

As Han et al. (2016), we assume that subhaloes follow the same spatial profile as the smooth DM component. In our case, this corresponds to a Hernquist (1990) profile~(\ref{eq:hern}), hence
\begin{align}\label{eq:g}
g(r)=\frac{1}{2\pi c_h^3}\frac{1}{(r/c_h)(1+r/c_h)^3}.
\end{align}
This assumption considerably simplifies the analysis below. For example, it allows us to derive the mean velocity dispersion of the subhalo population analytically from~(\ref{eq:W}) by setting $\nu=g$, which yields $\sigma_{\rm sub}^2=(1/18)GM_h/c_h$. An important consequence of our model is that the number density of subhaloes is separable in mass and radius. In practice, this implies that the probability of finding a subhalo within the mass bin $M,M+\d M$ remains uniform across the host dSph galaxy, further simplifying the statistical modelling outlined below.

According to Springel et al. (2008), the normalization of the subhalo mass function in the Aquarius simulations is $A_0=3.26\times 10^{-5}M_\odot^{-1}$ for a mass unit $M_0=2.52\times 10^7M_\odot$ in host haloes with a virial mass $M_{200}=1.839\times 10^{12}M_\odot$. Multiplying and dividing Equation~(\ref{eq:dNdM}) by $M_{200}^\alpha$, and integrating over a mass interval $(M_1,M_2)$ yields a total number of subhaloes
\begin{align}\label{eq:N}
N=\int_{M_1}^{M_2}\d M\frac{\d N}{\d M}=3.83\times 10^{-2}\bigg(\frac{1}{\xi_1^{0.9}}-\frac{1}{\xi_2^{0.9}}\bigg),
\end{align}
where $\xi\equiv M/M_{200}$, and $\xi_1\le \xi_2$. 
Assuming that the subhalo mass function is universal, Equation~(\ref{eq:N}) provides a simple estimate of the expected number of subhaloes with masses above $M_1$ in host galaxies of varying virial masses.

\begin{figure}
\begin{center}
\includegraphics[width=84mm]{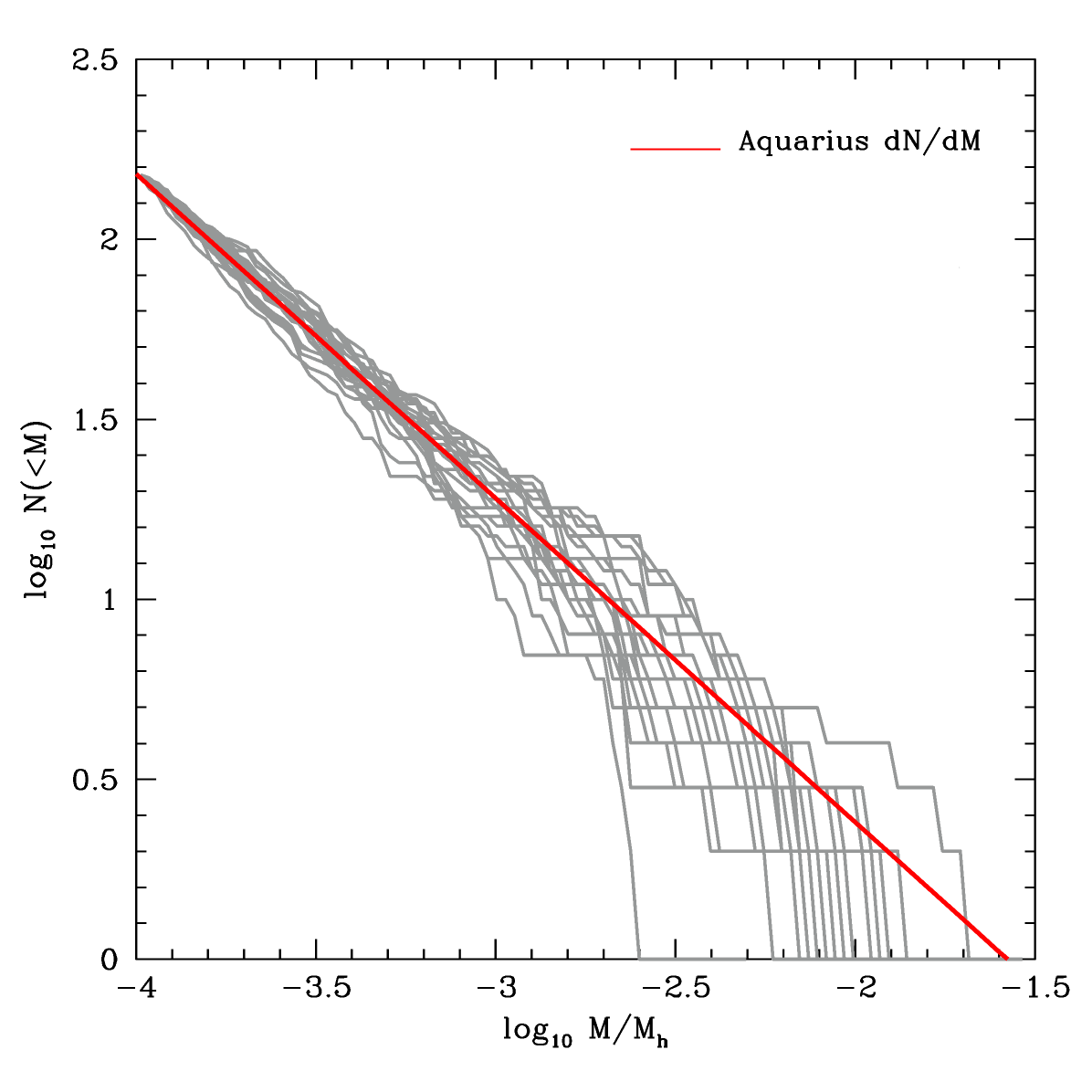}
\end{center}
\caption{Cumulative subhalo mass function sampled from 16 realizations of the Aquarius model (red line) within a mass range $\xi\equiv M/M_{200}>10^{-4}$. The total number of subhaloes derived from~(\ref{eq:N}) in each realization is $N\approx 152$.}
\label{fig:massfun}
\end{figure}

Fig.~\ref{fig:massfun} shows the cumulative mass function of 16 realizations of subhalo populations with $\xi>10^{-4}$, with the red line plotting the cumulative power-law function derived from~(\ref{eq:dNdM}). 
Importantly, relative to the host galaxy the vast majority of these objects have mass ratios $\xi\lesssim 10^{-2}$. This suggests that most subhaloes experience a negligible amount of dynamical friction (e.g. Pe\~narrubia \& Benson 2005). 
In this work, we construct subhalo ensembles with $\xi_1=10^{-4}$, and $\xi_1=10^{-5}$. For $\xi_2\gg \xi_1$, Equation~(\ref{eq:N}) can be approximated as $N(>M_1)\simeq 3.83\times 10^{-2}\,\xi_1^{-0.9}$, which yields an average number of subhaloes $N=151$ and $N=1208$, respectively.

Once the parameters $\{M_h,\xi_1,\xi_2\}$ have been specified, it is convenient to re-normalize the subhalo mass function~(\ref{eq:dNdM}) as $\d N/\d M=B_0(M/\msol)^{-\alpha}$, where
\begin{align}\label{eq:B0}
B_0=\frac{N\,(1-\alpha)}{\msol^\alpha(M_2^{1-\alpha}-M_1^{1-\alpha})},
\end{align}
with $B_0$ measured in units of $\msol^{-1}$ for convenience.

Next, we generate orbital ensembles in dynamical equilibrium within the halo potential. To this aim, we draw initial velocities from an
Opsikov-Merrit distribution function (Osipkov 1979; Merrit
1985). This returns orbital velocities with an anisotropy profile $\beta(r)\equiv 1-\overline{v^2_t}/(2\,\overline{v^2_r}) = r^2/(r^2 +r_a^2)$,
where $v_r$ and $v_t$ are radial and tangential velocity components, respectively, and $r_a$ is the so-called `anisotropy radius'.
Cosmological simulations show that the DM haloes of dSphs are radially anisotropic at large radii. The transition between isotropic and radially-anisotropic profiles occurs at a radius $r\sim 0.3\,c_h$ (He et al. 2024).
Motivated by cosmological models, we fix the anisotropy radius to $r_a=0.3\,c_h$, such that orbital velocities are isotropic, $\beta(r)\approx 0$, on scales populated by stars, $r\lesssim r_a=0.3\,c_h$, becoming radially-anisotropic, $\beta(r)>0$, at larger radii.

\begin{figure}
\begin{center}
\includegraphics[width=84mm]{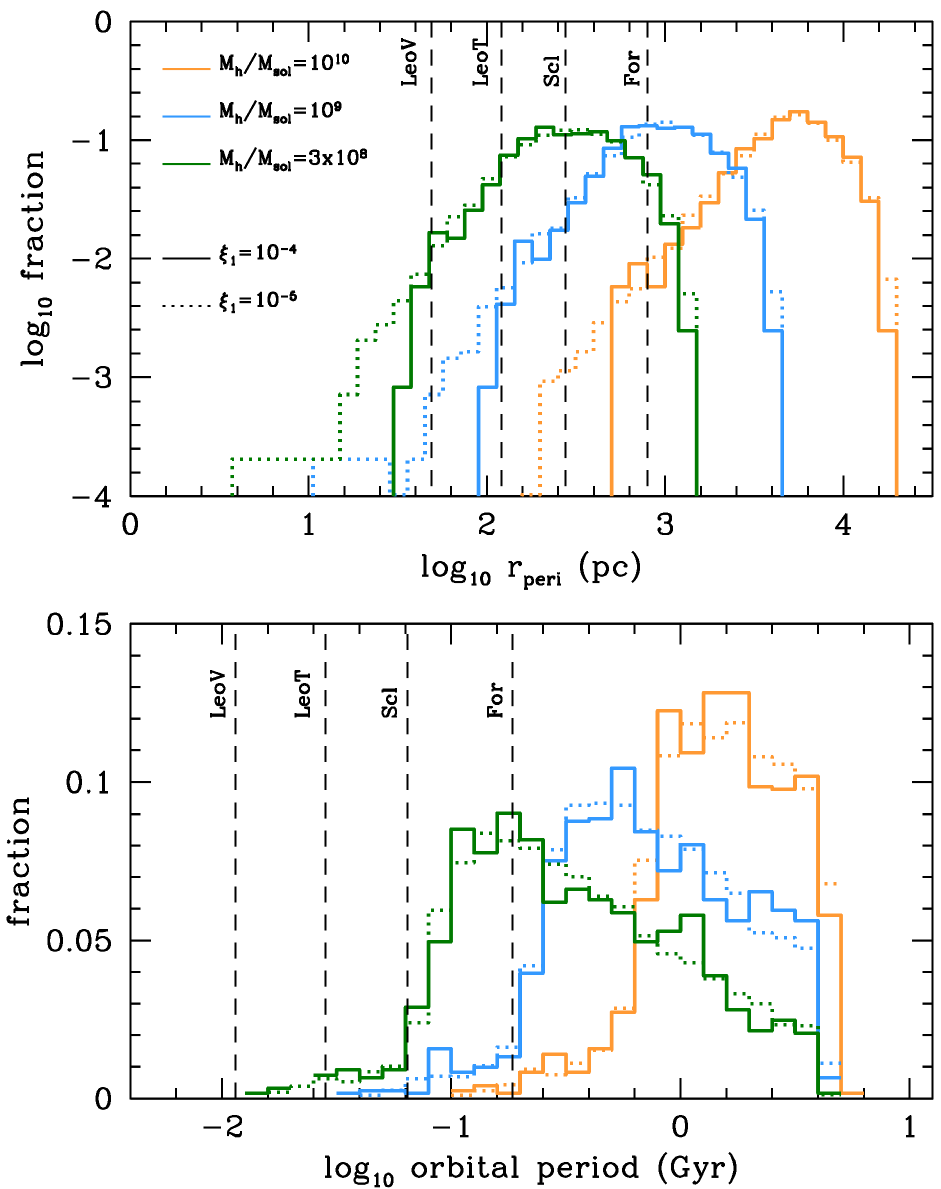}
\end{center}
\caption{Distribution of pericentric radii (upper panel) and orbital periods (lower panel) of subhaloes moving the in dSph haloes plotted in Fig.~\ref{fig:rhdens}. Solid and dotted lines show models with a low-mass truncation at $\xi_1=10^{-4}$ and $\xi_1=10^{-5}$, respectively. Black-dashed lines mark the half-light radii ($\rh$) and characteristic orbital times ($T_\star=2\pi \rh/\sigma$) of 4 dSphs used for reference in this paper. Notice that stars typically have shorter orbital periods than subhaloes because they are more spatially segregated in the halo potential.}
\label{fig:orb}
\end{figure}
\begin{figure}
\begin{center}
\includegraphics[width=80mm]{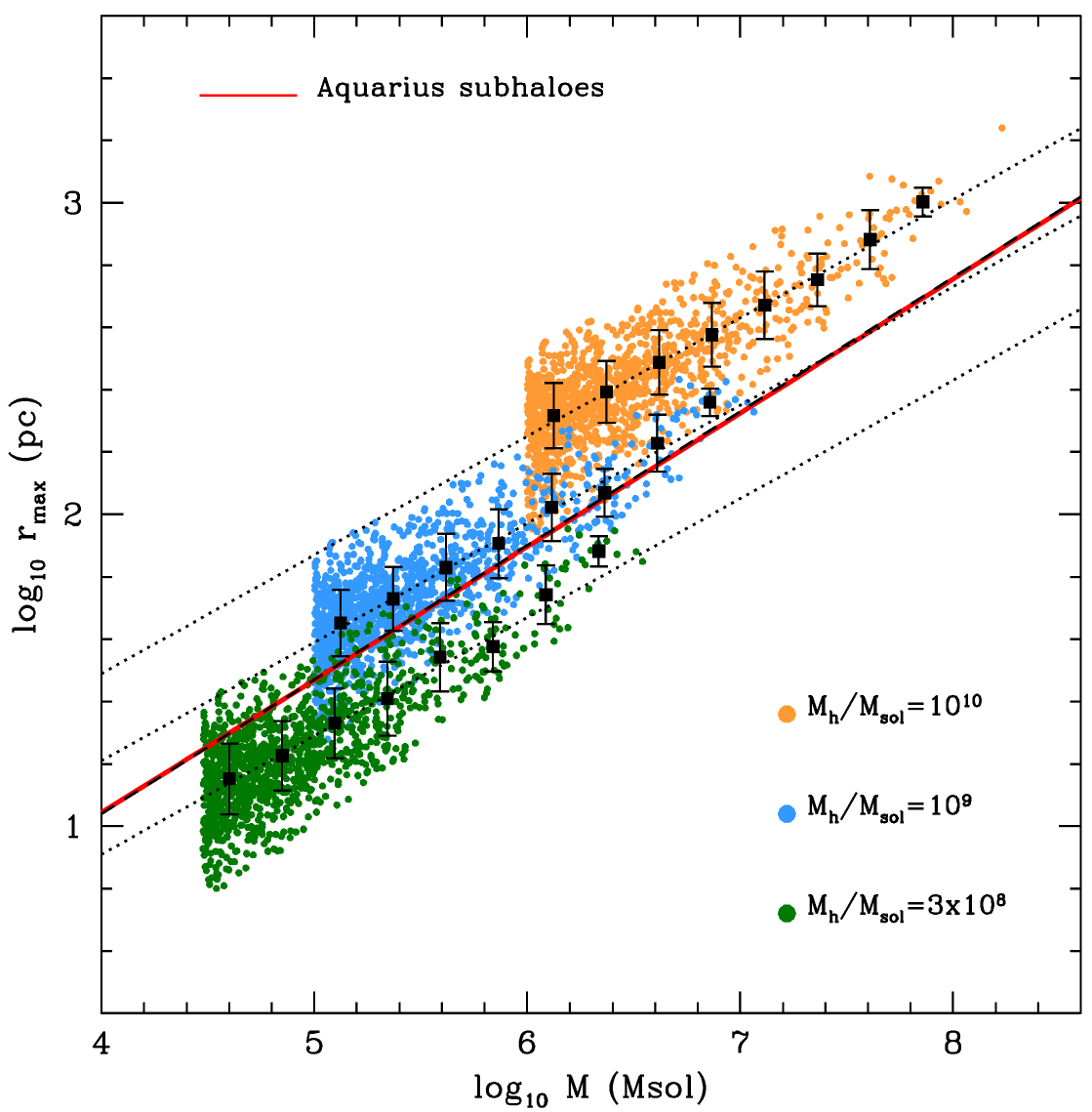}
\end{center}
\caption{Peak velocity radius of individual subhaloes with masses $10^{-4}<M/M_h<10^{-1}$ orbiting in the halo models plotted in Fig.~\ref{fig:rhdens}. Exponentially-truncated NFW profiles~(\ref{eq:rhoexp}) have a circular velocity profile that peaks at $r_{\rm max}\approx 1.8\,c$ (Errani \& Navarro 2021). Dotted lines show power-law fits $c=c_0(M/M_\odot)^\eta$ with $\eta=0.38$ and $c_0/\pc=0.25, 0.49$ and $0.93$ for dSph halo masses $M_h/M_\odot=3\times 10^8,10^9$ and $10^{10}$, respectively. For comparison, we overplot an extrapolation of the relation found in the Aquarius subhaloes by Springel et al. (2008), which exhibits a slightly steeper index $\eta=0.43$ with $c_0=0.21\pc$ (red line).}
\label{fig:rmaxm}
\end{figure}

\begin{figure*}
\begin{center}
\includegraphics[width=158mm]{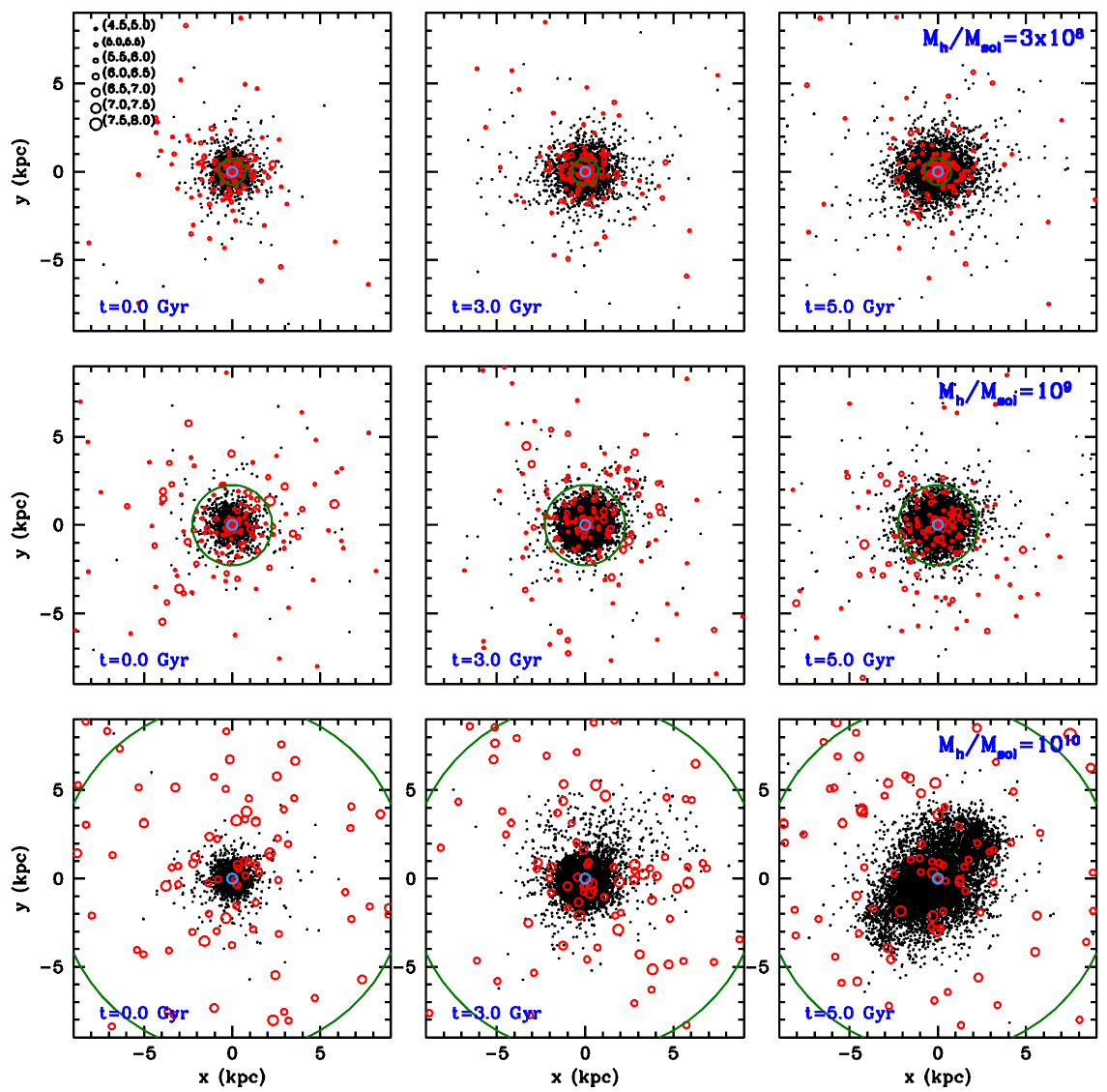}
\end{center}
\caption{Snap-shots of the evolution of a Sculptor-like dSph in the three halo models presented in \S\ref{sec:haloes} with $\xi_1=10^{-4} (N=151)$. Blue and green circles mark the size of the initial half-light radius, $\rh(t=0)=276\pc$, and the scale radius $c_h$ of the host DM halo, respectively. Subhaloes are shown with red open dots with a size that scales according to their mass, while black dots show stellar particles. Notice that as time goes by, dSphs tend to expand within the DM haloes as a result of repeated encounters with subhaloes, with signatures of disequilibrium that are more clearly visible in dSph halo models.}
\label{fig:xyz}
\end{figure*}

Subhaloes move on orbits dictated by the following equations of motion
 \begin{align}\label{eq:eqmotsun}
 \ddot {\bb r}_i=-\nabla\Phi_h({\bb r}_i),
 \end{align}
 where $i=1,..., N$, and $\Phi_h=-GM_h/(r+c_h)$ is the potential of the DM halo models plotted in Fig.~\ref{fig:rhdens}. Equation~(\ref{eq:eqmotsun}) neglects interactions between subhaloes, which are rare (e.g. Pe\~narrubia \& Benson 2005), as well as dynamical friction within the host halo. We discuss these approximations in \S\ref{sec:ufaint}.
 
The distribution of pericentric radii (upper panel) and orbital periods (lower panel) is shown in Fig.~\ref{fig:orb}. For reference, we mark with vertical dotted lines the half-light radii ($\rh$) and characteristic orbital times ($T=2\pi \rh/\sigma$) of 4 dSphs spanning a wide range of sizes and luminosities.

Fig.~\ref{fig:orb} reveals a number of interesting results. First, the distribution of orbital pericentres scales with the scale radius of the halo. As a result, most subhaloes orbiting in halo models with $M_h=10^{10}M_\odot$ have orbital pericentres that are much larger than the half-light radii of dSph. These objects move at large distances from stars and only rarely plummet into the inner-most regions of the galaxy. In contrast, in the lightest-DM halo ($M_h=3\times 10^{8}M_\odot$) a small fraction of subhaloes have orbital pericentres $r_{\rm peri}\lesssim 100\pc$, and may therefore experience close interactions with individual stars. 
Ultra-faint dSphs with $\rh \lesssim 50\pc$ provide an extreme case. Only subhaloes on very radial orbits will penetrate this region.
First, the abundance of subhaloes with very small pericentres is highly sensitive to the low-mass cutoff of the subhalo mass function. Comparing subhalo ensembles truncated at $\xi_1 = 10^{-4}$ (solid lines, with $N \simeq 150$ subhaloes) to those truncated at $\xi_1 = 10^{-5}$ (dotted lines, $N \simeq 1211$) reveals that limited resolution suppresses the sampling of orbits with small pericentres. This underscores the importance of resolution in accurately capturing the full range of dynamical interactions between stars and subhaloes. Second, even in the absence of direct close encounters, the cumulative gravitational influence of subhaloes can perturb stellar orbits.  Importantly, those perturbations cannot be formally considered `local'. We come back to this issue in \S\ref{sec:numerical}.

Another noteworthy result concerns the relative timescales of stellar and subhalo orbits. In general, the orbital periods of stars ($T_\star$) are shorter than those of the subhaloes, particularly in small galaxies. This disparity is most pronounced in ultra-faint dSphs, where stars orbit much more rapidly than subhaloes ($T_\star \ll T$), regardless of the host halo mass.

In more extended dwarfs, such as the Fornax dSph, the relationship between $T_\star$ and $T$ depends on the mass of the host halo. In low-mass haloes ($M_h = 3 \times 10^8~M_\odot$), stellar and subhalo orbital timescales are comparable ($T_\star \approx T$), whereas in more massive haloes ($M_h = 10^{10}~M_\odot$), the stars complete many orbits during a single subhalo passage ($T_\star \ll T$).

Fig.~\ref{fig:orb} also shows that a significant fraction of subhaloes follow orbits with relatively short periods ($T \lesssim 1$ Gyr), implying that many of them may complete multiple pericentric passages over a Hubble time. Repeated tidal interactions are expected to strip mass from subhaloes and reshape their internal structure. As shown by Errani \& Navarro (2021), the bound remnants of heavily stripped subhaloes tend to converge toward an NFW-like profile with an exponential truncation at large radii. Motivated by this result, we model subhaloes using the following density profile:
 \begin{align}\label{eq:rhoexp}
\rho(r)=\rho_0\bigg(\frac{c}{r}\bigg)\,\exp(-r/c).
 \end{align}
For a subhalo with a mass $M$ on an orbit with peri- and apocentres $r_{\rm peri}$ and $r_{\rm apo}$, respectively, the scale-length $c$ is determined by the mean density of the host at the pericentre (Errani \& Navarro 2021, Aguirre-Santaella et al. 2023). In particular, subhaloes are stripped gradually until their characteristic density $\langle \rho\rangle_{\rm ch} $ is approximately $16 \times$ the mean host density at pericentre, $\langle \rho_h\rangle_{\rm peri} \simeq 3M_h(<r_{\rm peri})/(4 \pi r_{\rm peri}^3)$. Hence, for each subhalo with a mass $M$ and a pericentre $r_{\rm peri}$, we derive the scale-length $c$ from~(\ref{eq:hern}) and~(\ref{eq:rhoexp}) by solving\footnote{Note that Errani \& Navarro (2021) define the characteristic density as $\langle \rho\rangle_{\rm ch} =3M(<r_{\rm max})/(4 \pi r_{\rm max}^3)$, where $r_{\rm max}\simeq 1.8\,c$ for truncated cusps. Using this definition returns $\sim 19\%$ smaller scale radii, which does not affect the results shown below.} the equation $\langle \rho\rangle_{\rm ch}=3 M/(4 \pi c^3)=16\,\langle \rho_h\rangle_{\rm peri}$. For ease of comparison with cosmological simulations, we also compute the radius where the circular velocity of the profile~(\ref{eq:rhoexp}) peaks, $r_{\rm max}=1.8\, c$ (Errani \& Navarro 2021). For consistency, we ensure that individual subhalo masses never exceed the host halo mass enclosed within the pericentric radius\footnote{These occurrences are rare, but when they happen we keep the mass and randomly redraw the subhalo orbit from the distribution function. }

Fig.~\ref{fig:rmaxm} shows the peak velocity radii of the subhalo populations orbiting the DM halo models plotted in Fig.~\ref{fig:rhdens} as a function of their individual masses. This plot shows several noteworthy results. First, independently of the host halo mass, the scale-radius--mass relation of subhaloes can be accurately described by a power-law function
 \begin{align}\label{eq:c}
c=c_0\bigg(\frac{M}{M_\odot}\bigg)^\eta,
 \end{align}
 with $\eta\approx 0.38$. The normalization $c_0$ increases slightly in dSph haloes with higher masses, in particular $c_0/\pc=0.25, 0.49$ and $0.93$ in dSphs with halo masses $M_h/M_\odot=3\times 10^8,10^9$ and $10^{10}$, respectively. Notice that these haloes have decreasing densities as a function of halo mass, $\rho_0=M_h/(2\pi c_h^3)=0.113, 0.0138$ and $ 0.0015/\msol/\pc^3$, respectively. Since the mass-size relation of subhaloes with a fixed orbital distribution is driven by the charaterisctic density of the host, the positive correlation between $c_0$ and $M_h$ in Fig.~\ref{fig:rmaxm} simply reflects the anticorrelation between $M_h$ and $\rho_0$ of the three dSph models considered in this work.
 Second, the scatter of the scale-radius--mass relation is $\sigma_{\log c}\approx 0.18$, with scant dependence on the subhalo masses, or the mass of the host dSph halo. This value is slightly larger than the scatter of the mass-concentration relation of field haloes. For example, Dutton \& Macci\`o (2014) find that $at z = 0$, the scatter around
the concentration–mass relation is well approximated by a
log-normal distribution of width $\sigma_{\log c_{\rm vir}}\simeq 0.11$ dex. Similarly, Molin\'e
et al. (2017, see their Fig. 4 and 5) measure a scatter between
0.10 -- 0.15 dex for the ELVIS (Garrison-Kimmel et al. 2014)
and VL-II (Diemand et al. 2008) simulations, whereas Nadler et al. (2023) find 0.12 dex. The slightly larger scatter in our models is not surprising given that the range of densities at a fixed mass increases for tidally-processed subhaloes (e.g. Pe\~narrubia et al. 2010a; Errani et al. 2022).
 Third, the scale-radius--mass relation of subhaloes in dSphs is similar to the curve extrapolated from subhaloes in the Aquarius simulation, albeit the latter has a slightly steeper index ($\eta\approx 0.43$). Again, this is result is to be expected given that the Aquarius halo has a charactersistic density comparable to that of our intermediate-mass dSph model. Indeed, $M_h=1.89\times 10^{12}\msol$ and $c_h=24.5\kpc$, which yields $\rho_0\simeq 0.02\msol/\pc^3$ (Springel et al. 2008).

\begin{figure*}
\begin{center}
\includegraphics[width=164mm]{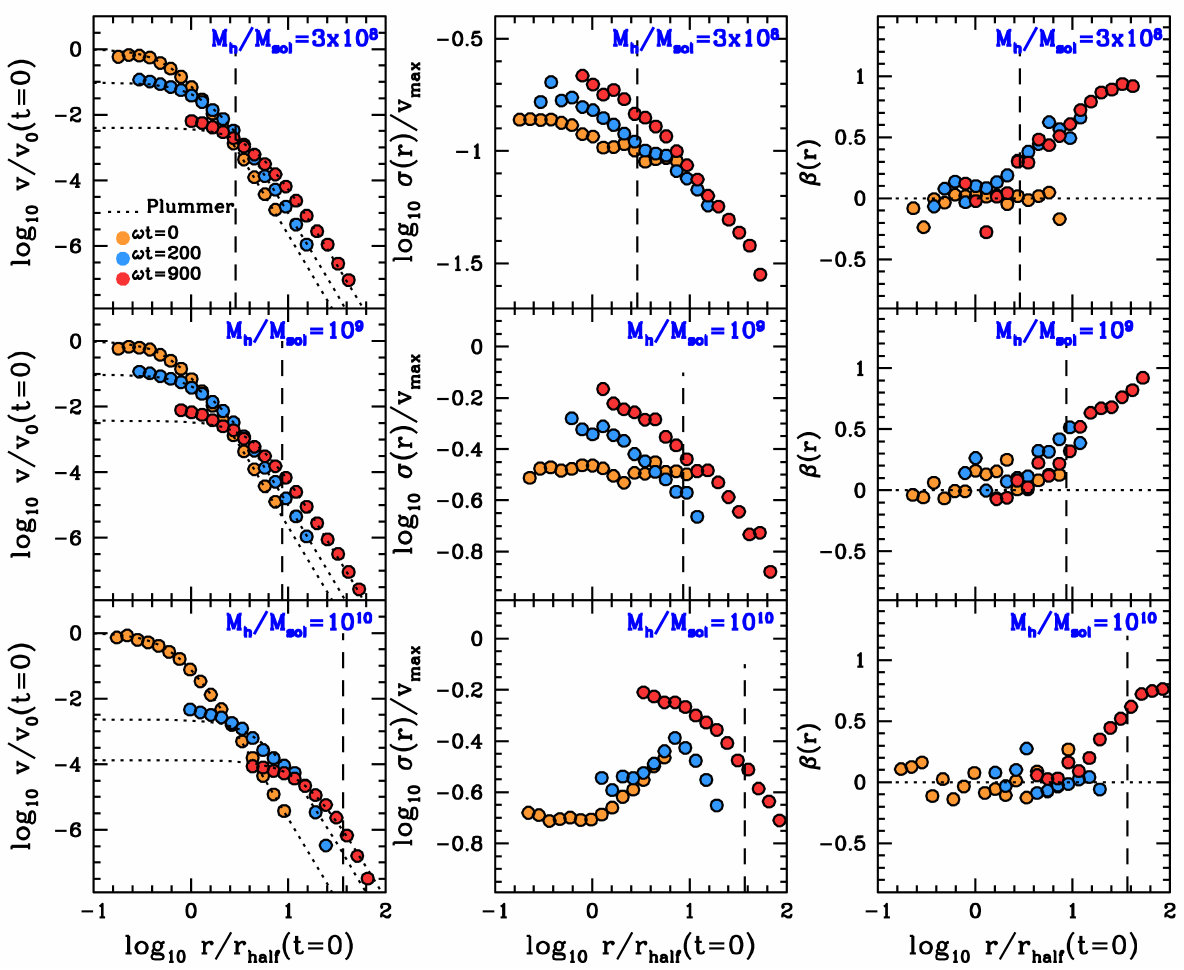}
\end{center}
\caption{{\it Left panels:} Density profile a Sculptor-like dSph in the three halo models presented in \S\ref{sec:haloes} with a low-mass truncation $\xi_1=10^{-4}$ at three different snapshots. Time is measured in units of the dynamical frequency of the dSph model at $t=0$, $\omega=\sigma/a$. The profiles are normalized by the central density measured at $t=0$, $\nu_0(t=0)$, while galactocentric radii are given in units of the initial half-light radius, $\rh(t=0)$. 
  Vertical dashed lines mark the location of the halo scale radius. Dotted-lines show the best-fitting Plummer profiles given by Equation~(\ref{eq:plummer}). Note that the stellar density profiles remain remarkably close to their initial Plummer distribution.
  {\it Middle panels:} Velocity dispersion profiles of the models plotted in the left panels, $\sigma^2(r)=(\sigma_\theta^2+\sigma_\phi^2+\sigma_r^2)/3$. Notice that as dSphs expand, stellar motions tend to heat up in the inner regions of the dwarf, $r\lesssim c_h$, whereas the velocity dispersion tends to drop at large radii.
{\it Right panels:} Velocity anisotropy profiles, $\beta(r)=1-(\sigma_\theta^2+\sigma_\phi^2)/(2\sigma_r^2)$. Stellar orbits that gain energy from the fluctuating force field become radially anisotropic at large radii.
}
\label{fig:distrib}
\end{figure*}

 \section{Internal evolution of dwarf spheroidals}\label{sec:evol}
In this section, we examine the dynamical evolution of stars embedded in dark matter (DM) haloes that host a large population of subhaloes. We begin by constructing initial stellar distributions using the Eddington inversion method,  which assumes that the initial velocity distribution is isotropic (see Errani \& Pe\~narrubia 2020). The stellar component follows a Plummer profile (Equation~\ref{eq:plummer}) and is in dynamical equilibrium within the DM halo (Equation~\ref{eq:hern}). Given the high mass-to-light ratios observed in dSphs (e.g., Simon et al. 2019), we treat stellar particles as massless tracers of the underlying gravitational potential.

The motion of a tracer star in a dSph galaxy is governed by the differential equation
 \begin{align}\label{eq:eqmot}
   \ddot {\bb r}_\star=-\nabla\Phi_h({\bb r}_\star)+\sum_{i=1}^N {\bb f}_i,
 \end{align}
 where ${\bb f}_i$ denotes the specific force induced by the $i$-th substructure on the stellar particle at a relative distance $({\bb r}_\star-{\bb r}_i)$. The subhalo trajectories ${\bb r}_i(t)$ are a solution to~(\ref{eq:eqmotsun}). Equations~(\ref{eq:eqmotsun}) and~(\ref{eq:eqmot}) are solved simultaneously for stars and subhaloes (for details of the numerical integration see Appendix A of Pe\~narrubia 2023).
 The two terms on the right-hand side of Equation~(\ref{eq:eqmot}) correspond to the forces exerted by the smooth and clumpy components of the DM halo, respectively. Note that the smooth component is modelled as a static halo potential, while subhaloes source static potentials and remain in dynamical equilibrium within the DM halo throughout our numerical experiments. These approximations are discussed in Section~\ref{sec:dis}.

 Figure~\ref{fig:xyz} presents three snapshots of a Sculptor-like dSph model evolving within three different halo models.
 The figure illustrates how stellar orbits {\it expand} in the presence of a fluctuating gravitational field. Despite this expansion, the galaxy approximately retains its initial spherical shape. However, the process is not smooth or continuous. Since stars predominantly occupy the central regions of the DM halo, the strongest perturbations arise from massive subhaloes passing through pericentre. Although such encounters are rare, they dominate the energy injected into stellar orbits.
 A particularly striking example is seen in the bottom-right panel of Fig.~\ref{fig:xyz}, where the dSph exhibits elongation and a transient shell-like structure near $(x,y) \approx (-3.5,-3.5)\kpc$. This feature is indicative of a recent interaction with a massive subhalo. Such disturbances typically persist for only a few dynamical times, after which the galaxy settles into a new equilibrium state with an increased half-light radius.

This evolution is illustrated in Figure~\ref{fig:distrib}, which shows the number density (left panels) and velocity dispersion (middle panels) profiles of the stellar component at three different snapshots. For convenience, these profiles are normalized by the initial central density, $\nu_0(t=0)$, and the radii are scaled by the initial half-light radius, $\rh(t=0)$, respectively. As the system expands, $\rh$ increases and the stellar density decreases at all radii. However, the number density can be approximately described by a Plummer profile (dotted lines) at all times, suggesting that the evolution is self-similar. This is the same behaviour observed in stellar tracer populations evolving within an adiabatically-evolving potential (Errani et al. 2025). Further experiments with different initial stellar profiles confirm that this self-similar evolution is independent of the initial conditions of the galaxy.

Over time, the velocity dispersion profile undergoes significant changes. Initially, the system has a nearly flat velocity dispersion profile, $\sigma(r)=[(\sigma_\theta^2+\sigma_\phi^2+\sigma_r^2)/3]^{1/2} \sim {\rm const.}$, but as the galaxy expands, the inner regions become kinematically hotter, while the outer regions cool down. This behavior is reminiscent of the evolution of stellar systems as a result of two-body encounters, which also leads to the formation of a hot core and a cool envelope. However, in this case, the process is mirrored: instead of collapsing, the galaxy undergoes expansion within a fluctuating dark matter potential.

The right panels of Figure~\ref{fig:distrib} show the evolution of the velocity anisotropy profile, $\beta(r)=1-(\sigma_\theta^2+\sigma_\phi^2)/(2\sigma_r^2)$. The initial stellar models have isotropic velocities, $\beta(r)\approx 0$ at all radii. However, as time progresses and stellar orbits gain energy, the system expands and the outer regions $r\gtrsim c_h$ develop a radially biased velocity distribution $\beta(r)>0$. 
Subhalo perturbations inject energy into stellar orbits, $\langle \Delta E\rangle >0$, but do change the mean angular momentum, $\langle \Delta {\bf L}\rangle =0$, leading to an increase in radial anisotropy at large radii.

\begin{figure}
\begin{center}
\includegraphics[width=80mm]{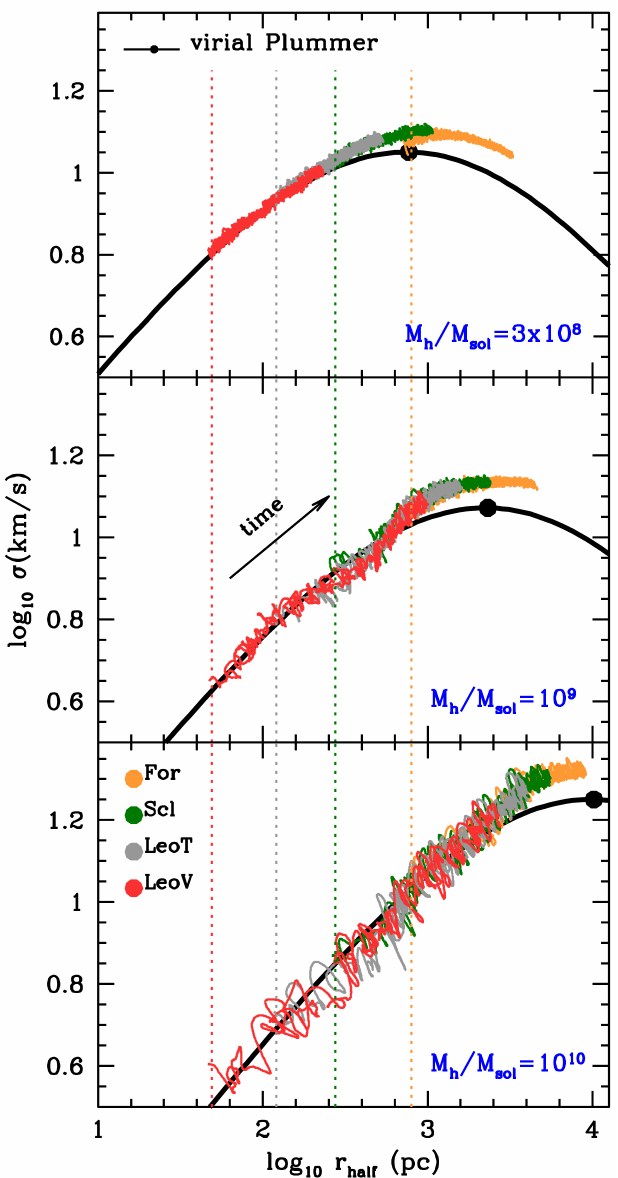}
\end{center}
\caption{Evolution in the half-light radius -- velocity dispersion plane of dSph models with different spatial segregations in the three halo models presented in \S\ref{sec:haloes} with $\xi_1=10^{-4} (N=151)$. Stellar and subhalo orbits are integrated for $t_{\rm final}=10^{3}\,\omega^{-1}$, where $\omega=\sigma/a$ is the dynamical frequency of the dSph measured at $t=0$. Lines are coloured-coded according to the initial half-light radius of the Plummer stellar component, $\rh=1.305\,a$. Vertical-dotted lines mark the initial half-light radii of the stellar components. Solid-black lines show the luminosity-averaged velocity dispersion of a Plummer stellar component embedded in a Hernquist DM halo $\sigma^2=-W$, where $W$ is the potential energy derived from Equation~(\ref{eq:W}).  Dots mark the location of the peak velocity radius. Notice that the size of dSphs expand and get kinematically hotter as a result of repeated encounters with subhaloes. As they evolve, dSphs remain close to virial equilibrium within the DM halo potential. }
\label{fig:rhsig}
\end{figure}

Fig.~\ref{fig:rhsig} exploits the self-similar behaviour of these models by following the evolution of the velocity dispersion as a function of half-light radius, $\rh(t)$, rather than time itself.
We find that the mean velocity dispersion of dSphs increases with time as the galaxy expands. This evolution follows closely the virial relation derived from Equation~(\ref{eq:W}), $\sigma^2=-W$, derived assuming a Plummer stellar profile embedded in a Hernquist DM halo. The fact that the numerical curves remain near the analytical prediction suggests that dSphs maintain quasi-virial equilibrium despite continuous perturbations.

Notably, the expansion of dSphs is more pronounced in galaxies that are initially more compact, and/or are embedded in massive DM haloes. For example, the final $\rh$ of ultra-faint dwarfs like Leo V has a strong dependence on the halo mass. After 900 dynamical times ($\omega\,t=900)$  $\rh$ increases by a factor of $\sim 4$ in a $M_h = 3 \times 10^8 \msol$ halo, whereas it grows by a factor of $\sim 50$ in a $M_h = 10^{10} \msol$ halo. In contrast, more extended systems like Fornax experience comparatively less growth, with $\rh$ increasing by a factor of $\sim 10$ in a $M_h = 10^{10} \msol$ halo.

As dSphs expand within the host DM potential, the velocity dispersion increases. Since these galaxies remain in a state of quasi-virial equilibrium at all times, we can use Equation~(\ref{eq:W}) to describe this process. First, it is straightforward to show that subhaloes with a number density profile~(\ref{eq:g}) have a potential energy $W_{\rm sub}=-(1/18)GM_h/c_h$. In contrast, the potential energy of compact galaxies with a Plummer profile~(\ref{eq:plummer}) can be expressed as $W\simeq -(2/3)GM_h\,a/c_h^2$ for $a/c_h\ll 1$. Hence, $\sigma^2/\sigma_{\rm sub}^2=W/W_{\rm sub}= 12\,(a/c_h)$, which shows that in the limit $a/c_h\to 0$ stellar orbits in the deepest regions of the host potential are `colder' than those of the subhaloes. However, as the stellar component expands, stars and subhaloes move with similar orbital velocities, $\sigma \sim \sigma_{\rm sub}$.

Importantly, the stellar component cannot become arbitrarily `hot'. The virial theorem constrains the maximum velocity dispersion that stars can reach. This occurs as the potential energy $W$ reaches a minimum at $a/c_h\simeq 0.76$ (or $\rh \approx c_h$). At this point of the evolution, the luminosity-averaged velocity dispersion peaks at $\sigma_{\rm max}^2=0.0735\,(GM_h/c_h)$, where $r_{\rm max}= c_h$ in the peak velocity radius of the host halo. In units of the peak circular velocity, this yields $\sigma_{\rm max}\approx 0.54\, v_{\rm max}$, where $v_{\rm max}=(1/2)\sqrt{GM_h/c_h}$ for a Hernquist sphere. Hence, the stellar velocity dispersion reaches its maximum ($\sigma_{\rm max}$) as the stellar half-light radius becomes comparable to the scale radius of the DM halo ($\rh\approx c_h$). This marks a fundamental transition in the evolution of dSphs, an issue we revisit in Section~\ref{sec:sto}.

In the next section, we use statistical tools to further investigate the relationship between galaxy expansion and the subhalo population.

\section{Dynamical heating}\label{sec:sto}
DM subhaloes orbiting around dSphs generate a combined force ${\bb F}=\sum {\bb f}_i$ that fluctuates randomly with time (e.g. Pe\~narrubia 2018). Stars acted on by a random force tend to absorb energy, which causes a progressive {\it expansion} of the size of dSph galaxies. This Section uses statistical tools to study the dynamical evolution of (mass-less) stellar tracers moving through a clumpy gravitational potential. For simplicty, the (smooth) halo potential is kept static. We discuss the effects of a time-varying potential in \S\ref{sec:dis}.

\begin{figure}
\begin{center}
\includegraphics[width=84mm]{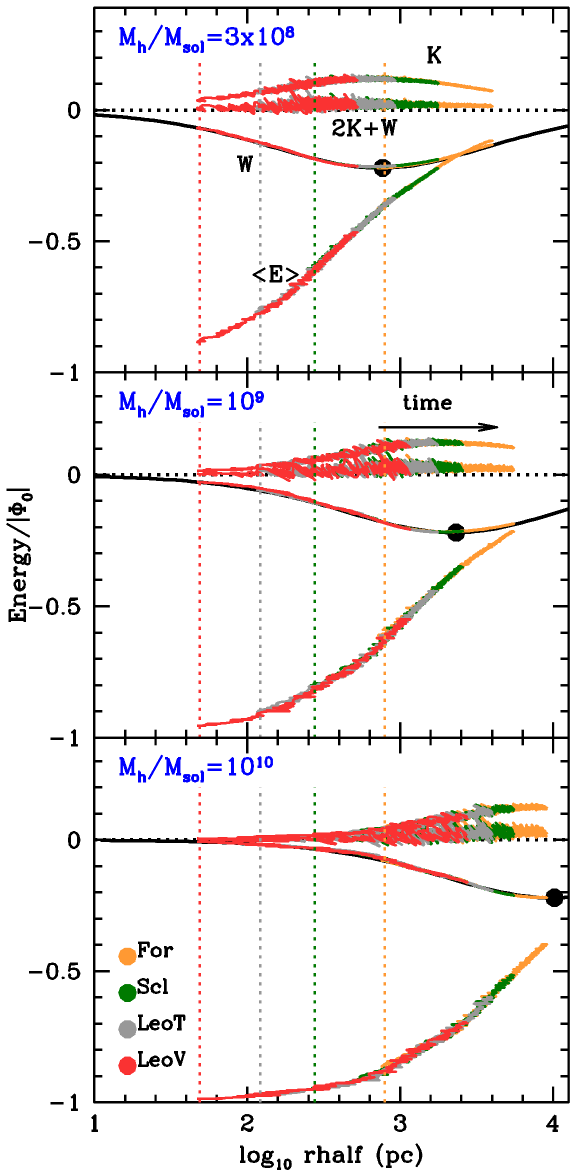}
\end{center}
\caption{Evolution of the mean kinetic energy ($K$), potential energy ($W$), virial energy ($2K+W$), and orbital energy ($\langle E\rangle$) as a function of half-light radius of the dSph models shown in Fig.~\ref{fig:rhsig}. Energies are normalized to the central halo potential, $\Phi_0=GM_h/c_h$. Lines are coloured-coded according to the initial half-light radius. Vertical-dotted lines mark the initial half-light radii of the stellar components. Solid-black lines show the potential energy of a Plummer stellar component embedded in a Hernquist DM halo~(\ref{eq:W}). Dots mark the location of the peak velocity radius. As expected, dSphs remain close to virial equilibrium ($2K+W\approx 0$) during early phases of the evolution, with disequilibrium ($2K+W\gtrsim 0$) progressively arising as they expand. Notice that the slope of the kinetic (and potential) energy vanishes close to the peak velocity radius, which marks a change of sign of the heat capacity of the galaxy (see text). In contrast, the mean orbital energy increases monotonically with time, which indicates that stars become progressively less bound.}
\label{fig:W}
\end{figure}

\subsection{Virial equations}\label{sec:vir}
As a first step, we aim to determine whether dSphs remain in a state of quasi-virial equilibrium during their expansion within the host potential. To this end, we compute the mean kinetic energy, $K=(1/2)\sum_{i=1}^{N_\star}v_i^2/N_{\star}$ and the mean potential energy $W=-\sum_{i=1}^{N_\star}r_i|\nabla\Phi_i|/N_{\star}$ of the stellar component. Additionally, we track the average orbital energy of individual stars, defined as $\langle E\rangle =\sum_{i=1}^{N_\star}E_i/N_\star=K + \sum_{i=1}^{N_\star}\Phi_i/N_\star$. 

Figure~\ref{fig:W} illustrates the evolution of the virial energy components for the dSph models presented in Fig.~\ref{fig:rhsig}. Since the half-light radius, $\rh$, increases monotonically with time, models evolve from left to right in this diagram. Several noteworthy results emerge from this analysis. First, $K$ and $W$ evolve in such a way that the virial theorem $2 K+W\approx 0$ is approximately satisfied in the early stages, indicating that these dSphs are initially close to virial equilibrium. However, as time progresses, the virial sum becomes weakly positive, $2 K+W \gtrsim 0$, which is a clear sign that these galaxies are drifting out of equilibrium. 

A second key result is that dSphs reach their maximum kinetic energy and minimum potential energy precisely when their half-light radius matches the scale radius of the DM halo, $\rh = c_h$. 
This can be understood by inspecting Equation~(\ref{eq:W}). To gain physical intuition, consider the case where all stars are located within a shell at a radius $r=r_\star$. Accordingly, the stellar number density can be written as a delta function $\nu(r)=(4\pi r_\star^2)^{-1}\delta(r-r_\star)$, with a normalization chosen such that $\int \d^3 \nu(r)=1$. Equation~(\ref{eq:W}) can be trivially integrated, which yields $\sigma^2=(1/3)GM_h(<r_\star)/r_\star$. 
Notice that the radial behaviour of the velocity dispersion depends on the spatial segration of the stellar component within the DM halo. 
Stars deeply embedded in the potential have a velocity dispersion that grows linearly as $\sigma^2=(1/3)GM_h \,r_\star/c_h^2\sim r_\star$ for $r_\star\ll c_h=r_{\rm max}$. However, once the size of the stellar components exceeds the peak velocity radius of the halo, the stellar dispersion declines as $\sigma^2=(1/3)GM_h/r_\star\sim r_\star^{-1}$ for $r_\star\gtrsim c_h=r_{\rm max}$. Accordingly, the mean kinetic energy of the stellar particles ($K$) decreases at large radii, as observed in Fig.~\ref{fig:rhsig}, reflecting how the potential energy of the host DM halo varies with radius.


In contrast, the average orbital energy $\langle E\rangle$ grows monotonically with time. At early times, it lies below the potential energy, $E<W$. This is because in the central regions of the halo the integrand of Equation~(\ref{eq:W}) scales as $GM(<r)/r\sim r/(r+c_h)^2\to 0$, whereas the halo potential approaches its lowest (most bound) value, $\Phi_h(r)\to -GM_h/c_h$. As galaxies expand, the average orbital energy converges towards their potential energy, such that $\langle E\rangle\approx W$ at $\rh\gtrsim c_h$.

Notice also that galaxies that are deeply embedded in the DM potential ($\rh\ll c_h$) become kinematically {\it hotter} as they expand. Thermodynamically, this implies stellar tracers have {\it positive} heat capacity,  $c\equiv \d \langle E\rangle /\d K$, in stark contrast with self-gravitating systems, which have negative heat capacities (e.g. Lynden-Bell 1999 and references therein). 
However, this trend reverses as the half-light radius exceeds the peak velocity radius, $\rh\gtrsim r_{\rm max}=c_h$. At this point, stellar orbits begin to {\it cool down}, and the the sign of the specific heat of the galaxy flips from positive to negative, significantly complicating any attempt to derive the growth rate of the galaxy ($\d \rh/\d t$) using virial quantites (see for example Brandt 2016).

\begin{figure}
\begin{center}
\includegraphics[width=76mm]{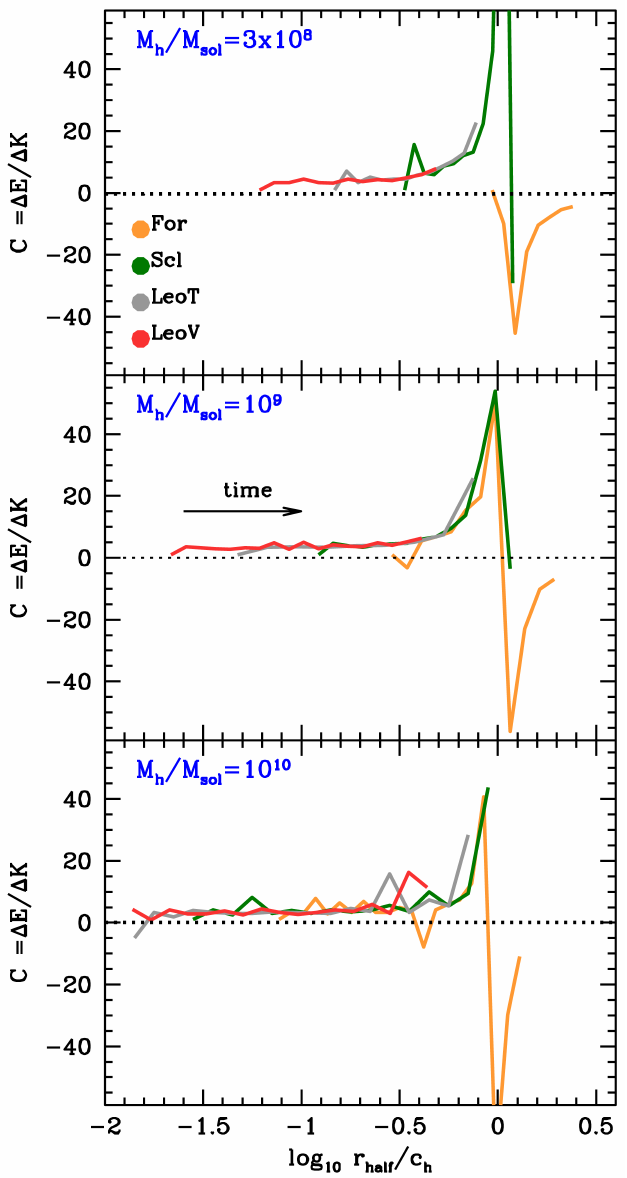}
\end{center}
\caption{ Heat capacity as a function of the half-light radius of the dSph models shown in Fig.~\ref{fig:W}.
  Stellar tracers deeply embedded in the dark matter potential exhibit a positive heat capacity, meaning their mean kinetic energy (a proxy for gravitational temperature) increases as energy is injected into their orbits. As the half-light radius approaches the scale radius of the dark matter halo ($\rh\approx c_h$), the heat capacity diverges, resembling a second-order phase transition where a system undergoes a critical change in response to energy injection. Beyond this threshold ($\rh>c_h$), the heat capacity becomes negative, indicating that additional energy input leads to a decrease in velocity dispersion. Notably, these models follow a self-similar evolution governed by the embedness ratio $\rh/c_h$.}
\label{fig:cv}
\end{figure}

Figure~\ref{fig:cv} illustrates this transition, showing that when $\rh \ll c_h$, dSphs exhibit positive heat capacity, meaning that increasing energy input leads to higher velocity dispersions. As $\rh$ approaches $c_h$, the heat capacity diverges, resembling a second-order phase transition in which the system undergoes a critical structural shift. Beyond this point ($\rh > c_h$), the heat capacity turns negative, meaning that additional energy input causes velocity dispersion to decrease. 
This behavior highlights the self-similar nature of the expansion, which is primarily governed by the ratio $\rh / c_h$.

Importantly, Fig.~\ref{fig:W} thus reveals that potential fluctuations generated by subhaloes act as a {\it heat source}, and that stars react to those fluctuations by becoming progressively less bound within the DM potential well, i.e. $\langle \Delta E\rangle>0$. We study this process in detail in \S\ref{sec:kinetic}.

\subsection{Kinetic theory}\label{sec:kinetic}
Here we abandon the thermodynamical approach and turn instead to kinetic theory to understand the time-variation of the orbital energy of {\it individual} stellar particles, $E_i=v_i^2/2+\Phi_h(\bb r_i)$ with $i=1,\dots, N_\star$.

\subsubsection{Random velocity kicks}\label{sec:kicks}
As a first step, we follow up the pioneer work of Chandrasekhar (1941b), who was the first to consider the effect on stellar dynamics of a noise term in right-hand term of Equation~(\ref{eq:eqmot}). In his theory, Chandrasekhar computes the probability $p({\mathbfit F})$ that a tracer star moving in an homogeneous sea of point-masses with a constant number density $n$ experiences a combined force ${\bb F}=\sum_{i=1}^N {\bb f}_i$ within the interval ${\mathbfit F},{\mathbfit F}+\d{\bb F}$. From the spectrum of potential fluctuations, Chandrasekhar calculates the average squared velocity impulse $\langle |\Delta {\bb v}|^2\rangle$ that would result from a large number of force fluctuations within a short time interval $t$.

Pe\~narrubia (2019b; hereafter P19b) extended Chandrasekhar's derivation to perturbers with an extended Hernquist (1990) density profile. The behaviour of $\langle |\Delta {\bb v}|^2\rangle$ was found to be very sensitive to the time-scale ($T$) on which the force $\bb F$ fluctuates. According to Smoluchowski (1916), the fluctuation mean-life is set by the duration of a state in which the number of substructures within a small volume $V=4\pi r^3/3$ centred around a star remains constant
\begin{align}\label{eq:TR}
  T(r)={\sqrt\frac{2\pi}{3 \langle v^2\rangle} }\frac{r}{\frac{4\pi }{3}r^3n+1},
\end{align}
where $n=N/V$ is the number density of substructures distributed homogeneously around a star, and $\langle v^2\rangle$ is the mean {\it relative} velocity between the star and the surrounding substructures.
Within a small volume, $4\pi r^3 n/3\ll 1$, Equation~(\ref{eq:TR}) approaches the time that a straight-line trajectory takes to cross the radius $r$, i.e. $T(r)\sim r/\sqrt{\langle v^2\rangle}$, which vanishes in the limit $r\to 0$.
At large distances $4\pi r^3 n/3 \gg 1$, the probability that a substructure leaves/enters $V$ becomes proportional to the number of substructures within this volume. Hence, Equation~(\ref{eq:TR}) scales as $T\sim r^{-2}$, which vanishes in the limit $r\to \infty$. It is straightforward to show that $T(r)$ peaks at 
\begin{align}\label{eq:T0} 
 T_0={\sqrt\frac{2\pi}{3} }\frac{D}{\langle v^2\rangle^{1/2}},
\end{align}
where $D=(2\pi n)^{-1}$ is the average distance of the closest subhalo. Thus, Equation~(\ref{eq:T0}) sets the average number of fluctuations that a star experiences during a time-interval $t$.

P19b finds that on long time scales, $t\gg T_0$, the variance of the velocity impulses generated by a large number of extended substructures with a mass $M$ and a scale radius $c$ is
\begin{align}\label{eq:delv2_long}
  \langle |\Delta{\mathbfit v}|^2\rangle_d =t\,\sqrt{\frac{32\pi^3}{3\langle v^2\rangle}}(GM)^2n\big[\ln (D/c) -1.9\big] ~~~~~{\rm for}~~~~t\gg T_0.
\end{align}
In this so-called `dynamic' limit, the system exhibits a typical stochastic behaviour, with a variance that increases in proportion to the length of the time-interval, $\langle|\Delta{\mathbfit v}|^2 \rangle\propto t$.
As originally pointed out by Chandrasekhar (1941a,b), the `Coulomb logarithm' $\ln (\Lambda)\equiv \ln(D/c)$, diverges in the point-mass limit $c/D\to 0$. Using numerical experiments, P19b found that the median\footnote{Note that P19b also shows that the {\it average} of $\ln(\Lambda)$ diverges in the limit $c/D\to 0$, even though the median has a finite value.} value of the Coulomb logarithm flattens at a maximum value $\langle \ln(\Lambda)\rangle \approx 8.2$ for $c/D\lesssim 10^{-3}$ (see Fig. 3 of P19b).

In contrast, over very short time intervals, $t\ll T_0$, the relative position between stars and the surrounding substructures will not change appreciably with time. In this so-called `static' regime, P19b finds that the averaged squared velocity impulses increases as
\begin{align}\label{eq:delv2_short}
    \langle|\Delta{\mathbfit v}|^2\rangle_s  \simeq t^2\,55.8 \,(GM)^2 n^{4/3}  ~~~~~{\rm for}~~~~t\ll T_0,
\end{align}
which has a relatively fast growth, $\langle|\Delta{\mathbfit v}|^2\rangle\sim t^2$. Notice that in this regime velocity impulses do not depend on the average velocity between stars and subhaloes ($\langle v^2\rangle$) or the subhalo scale-radius ($c$).

Below, it is helpful to interpolate between the {\it static} and {\it dynamic} regimes using a linear relation
\begin{align}\label{eq:delv2_int}
  \langle|\Delta{\mathbfit v}|^2\rangle= \xi\,\langle|\Delta{\mathbfit v}|^2\rangle_s + (1-\xi)\langle|\Delta{\mathbfit v}|^2\rangle_d,
\end{align}
where the coefficients $\langle|\Delta{\mathbfit v}|^2\rangle_d$ and $\langle|\Delta{\mathbfit v}|^2\rangle_s$ are given by Equations~(\ref{eq:delv2_long}) and~(\ref{eq:delv2_short}), respectively. P19b finds that the transition between the two asymptotic behaviours happens on a time-scale comparable to the fluctuation mean-life~(\ref{eq:T0}), and that a linear interpolation, $\xi=1/(1+x)$ with $x=t/(T_0/2)$, reproduces reasonably well the results of numerical experiments.

It is important to emphasize that neither 
Equation~(\ref{eq:delv2_long}) nor~(\ref{eq:delv2_short}) depend on the substructure scale-radii insofar as the separation between substructures is much larger than their individual sizes, $D\gg c$.
This implies that the heating process has scant sensitivity to the subhalo mass-size relation plotted in Fig.~\ref{fig:rmaxm}.

\begin{figure*}
\begin{center}
\includegraphics[width=156mm]{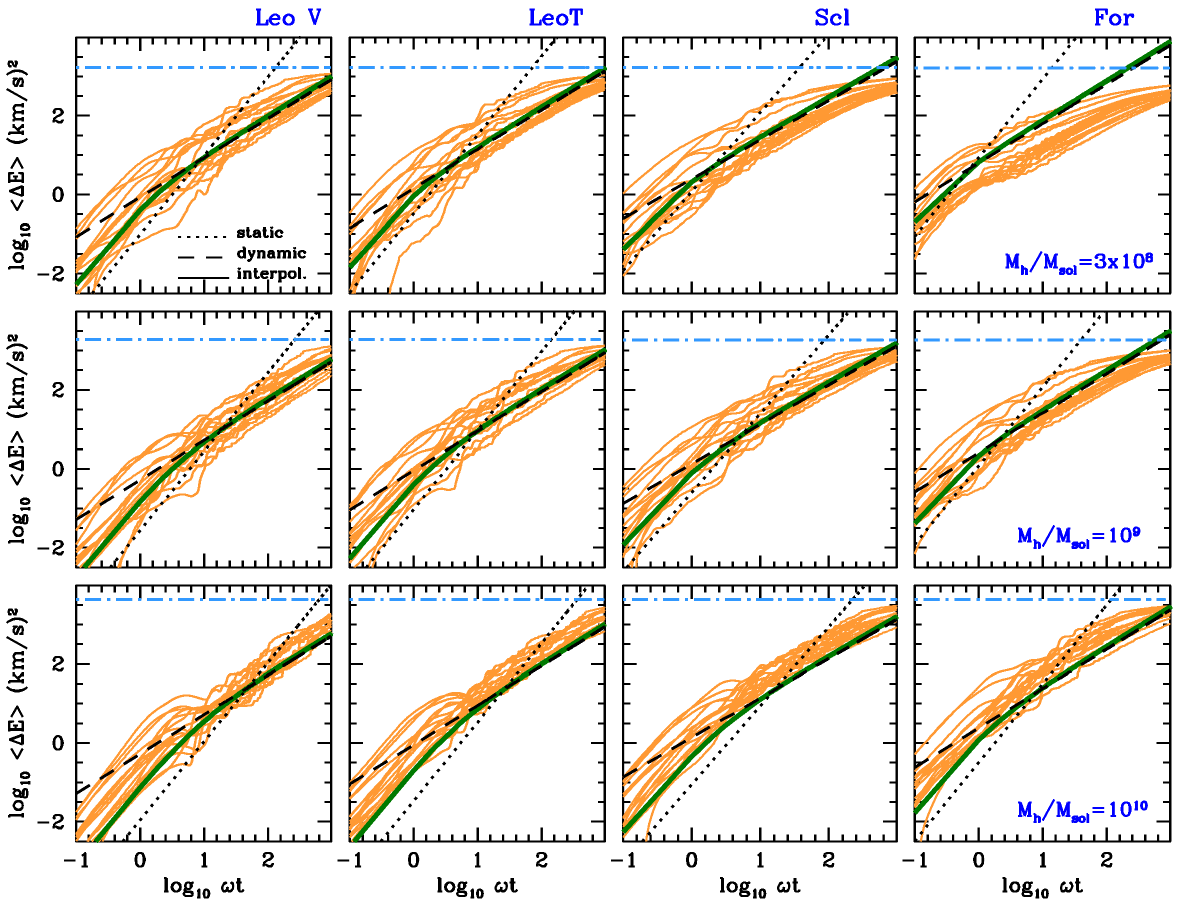}
\end{center}
\caption{Time-variation of the mean orbital energy ($\langle \Delta E\rangle$) of the dSph models shown in Fig.~\ref{fig:W}. Individual lines correspond to random realizations of a subhalo population. Green-solid lines show the interpolation between the static (dotted lines) and dynamic (dashed lines) regimes. Blue dotted-dashed lines show the maximum energy that the stellar component can absorb via encounters with subhaloes, $\langle \Delta E\rangle_{\rm max}\approx GM_h/c_h$ (see text).}
\label{fig:de10}
\end{figure*}
\subsubsection{Heating by DM subhaloes}\label{sec:heating}
The presence of a noise term in the equations of motion~(\ref{eq:eqmot}) means that the integrals $\{E,{\mathbfit L}\}$ are not conserved quantities. Following Pe\~narrubia (2019a; hereafter P19a), we use the average variance of velocity impulses to estimate the amount of heating introduced to stellar motions as a result of force fluctuations generated by subhaloes.

In the impulse regime, the location of a tracer particle is assumed to remain constant. Hence, the variation of orbital energy is equal to the change of kinetic energy
\begin{align}\label{eq:delE}
   \Delta E=\frac{1}{2}({\mathbfit v}+\Delta {\mathbfit v})^2-\frac{1}{2}{\mathbfit v}^2={\mathbfit v}\cdot\Delta{\mathbfit v}+\frac{1}{2}(\Delta {\mathbfit v})^2,
\end{align}
while the angular momentum varies by an amount
\begin{align}\label{eq:delL}
  \Delta {\mathbfit L}={\mathbfit r}\times({\mathbfit v}+\Delta {\mathbfit v})- {\mathbfit r}\times{\mathbfit v}={\mathbfit r}\times\Delta {\mathbfit v}.
\end{align}
Assuming that the population of substructures are isotropically distributed around stars, the average variation of the integrals becomes
\begin{align}\label{eq:delEL}
  \langle \Delta E\rangle &=\langle{\mathbfit v}\cdot\Delta{\mathbfit v}\rangle+ \frac{1}{2}\langle|\Delta {\mathbfit v}|^2\rangle=\frac{1}{2}\langle|\Delta {\mathbfit v}|^2\rangle \\ \nonumber 
   \langle \Delta {\mathbfit L}\rangle &=\langle{\mathbfit r}\times\Delta {\mathbfit v}\rangle=0,
\end{align}
which shows that isotropic force fluctuations tend to inject energy to stellar orbits, $\langle \Delta E\rangle>0$, while keeping the total angular momentum invariant, $\langle \Delta {\mathbfit L}\rangle =0$. This largely agrees with the mean energy evolution plotted in Fig.~\ref{fig:W}.


The above results can be straightforwardly extended to ensembles of substructures with mass function~(\ref{eq:dndm}) by writing the number density as
$n\to \frac{\d n}{\d M}\d M$, and integrating over a mass interval $M\in (M_1,M_2)$.

On long time-scales $t\gtrsim T_0$, the average velocity impulses~(\ref{eq:delv2_long}) becomes
\begin{align}\label{eq:delv2_long_dndm}
  \langle |\Delta{\mathbfit v}|^2\rangle_d =t\,B_0\,g\sqrt{\frac{32\pi^3}{3\langle v^2\rangle}}\frac{G^2\msol^\alpha}{3-\alpha}\bigg(M_2^{3-\alpha}-M_1^{3-\alpha}\bigg)(\ln \Lambda -1.9),
\end{align}
where we have assumed that subhaloes have sizes much smaller than their average separation ($c/D\ll 1$), such that the Coulomb logarithm can be approximated as a constant, $\ln (D/c)\approx \ln (\Lambda)\approx 8.2$. 

On short time-intervals $t\lesssim T_0$ Equation~(\ref{eq:delv2_short}) becomes
\begin{align}\label{eq:delv2_short_dndm}
  \langle |\Delta{\mathbfit v}|^2\rangle_s =t^2\,30.2\,B_0\,g\,\frac{G^2\msol^\alpha}{D(3-\alpha)}\bigg(M_2^{3-\alpha}-M_1^{3-\alpha}\bigg),
\end{align}
To derive an analytical expression for~(\ref{eq:delv2_short_dndm}) we have written $n^{4/3}=n\,n^{1/3}=n/[(2\pi)^{1/3}D]$, where $D$ is the average separation between subhaloes located at a galactocentric distance $r$ 
\begin{align}\label{eq:D_dndm}
  D&=\bigg[2\pi \int_{M_1}^{M_2} \frac{\d n}{\d M}\d M\bigg]^{-1/3} \\ \nonumber
  &=\bigg[2\pi\,B_0\,g\frac{\msol^\alpha}{1-\alpha}\bigg(M_2^{1-\alpha}-M_1^{1-\alpha}\bigg)\bigg]^{-1/3}.
\end{align}
Although this is a crude average, we will see below that it provides a reasonable description of the numerical results.

Let us now apply the above equations to CDM haloes. According to cosmological simulations of structure formation, subhaloes follow a mass function with a steep power-law index $\alpha\approx -1.9$ (e.g. Springel et al. 2008), with the smallest subhaloes having masses that lie many orders of magnitude below the highest subhalo mass, $M_1\lll M_2$. This implies that the most abundant subhaloes in a DM halo have masses $M\sim M_1$, and are separated by an average distance
\begin{align}\label{eq:D_CDM}
  D_{\rm CDM}\approx \bigg[2\pi\,B_0\,g\frac{\msol^\alpha}{\alpha-1}M_1^{1-\alpha}\bigg]^{-1/3},
\end{align}
which scales as $D_{\rm CDM}\sim M_1^{0.3}$ for $\alpha=1.9$.

In contrast, on long time-scales the coefficients~(\ref{eq:delv2_long_dndm}) and~(\ref{eq:delv2_short_dndm}) are dominated by the largest subhaloes with $M\sim M_2$, hence
\begin{align}\label{eq:delv2_long_CDM}
  \langle |\Delta{\mathbfit v}|^2\rangle_{\rm CDM} \approx B_0\,g\frac{G^2\msol^\alpha}{3-\alpha}M_2^{3-\alpha} \times
  \begin{cases}
    t\,\sqrt{\frac{32\pi^3}{3\langle v^2\rangle}}(\ln \Lambda -1.9)&, t\gtrsim T_0\\
    t^2\, 30.2 /D_{\rm CDM} &, t\lesssim T_0,
    \end{cases}
\end{align}
A key result from Equation~(\ref{eq:delv2_long_CDM}) is that lowering the minimum subhalo mass $M_1\to 0$ ($D\to 0$) while keeping the normalization of the mass function $B_0$ fixed systematically enhances the amplitude of velocity impulses on short time-intervals ($t\lesssim T_0$), but does not affect the long-term evolution of the stellar system ($t\gtrsim T_0$). This limits the information on the low-end of the subhalo mass function that can be extracted from the structural properties of dSphs.

\subsubsection{Numerical results}\label{sec:numerical}
To test the analytical equations derived above, Fig.~\ref{fig:de10} plots the mean orbital energy $\langle \Delta E\rangle$ as a function of time of stellar tracers orbiting in the dSph models discussed in Fig.~\ref{fig:rhdens}, with each individual line representing a random realization of the subhalo population.
Dashed and dotted-black lines show the analytical expression $\langle \Delta E\rangle=\langle|\Delta {\mathbfit v}|^2\rangle/2$ derived from~(\ref{eq:delv2_long_dndm}) and~(\ref{eq:delv2_short_dndm}), respectively, while the green-solid line corresponds to the interpolation between the static and dynamic regimes~(\ref{eq:delv2_int}).
All analytical coefficients are computed at $t=0$. Following Fig.~\ref{fig:massfun}, we set the limits of the mass function to $(\xi_1,\xi_2)=(10^{-4},0.03)$, whereas the mean velocity between stars and subhaloes is calculated as $\langle v^2\rangle = 3\sigma_{\rm sub}^2+3\sigma_0^2$, with $\sigma_0\equiv \sigma(t=0)$. Since subhaloes and DM particles follow the same spatial profile, we use the mean velocity dispersion of a Hernquist sphere, $\sigma_{\rm sub}^2=(1/18)GM_h/c_h$.

Estimating the subhalo number density is more challenging. In principle, the mass function in Eq.~(\ref{eq:dndm}) is separable in mass and radius, implying that the spatial distribution of subhaloes should be independent of their mass. However, this separability only holds asymptotically in the limit of large subhalo numbers ($N \to \infty$). In our simulations, the number of subhaloes is finite, and the distribution function is therefore sampled coarsely. As a result, the number of subhaloes found at small galactocentric distances at any given time is limited by resolution.
This limitation is illustrated in Fig.~\ref{fig:orb}, which shows that subhalo ensembles truncated at $\xi_1=10^{-4}$ contain no orbits with pericentres $r_{\rm peri} \lesssim 0.1\,c_h$. Reducing the minimum subhalo mass to $\xi_1=10^{-5}$ increases the total number of subhaloes by a factor $\sim 10$ and enables sampling of smaller pericentres, but still fails to produce any orbits with $r_{\rm peri} \lesssim 0.05\,c_h$.
Consequently, at radii below the smallest pericentre in the orbital sample the subhalo number density effectively vanishes ($g \approx 0$), which is at odds with the divergent behaviour ($g \sim r^{-1}$) expected in the $N \to \infty$ limit. This suppression must be accounted for in Eqs.(\ref{eq:delv2_long_dndm}) and(\ref{eq:delv2_short_dndm}); otherwise, those expressions would significantly overestimate the heating of stellar orbits in compact galaxies with $\rh \lesssim 0.1\,c_h$.
In the analysis below, we correct for this effect by assuming a constant subhalo number density equal to $g(c_h)$, which provides a reasonable approximation for the range models explored in this work.

As expected from Equation~(\ref{eq:delv2_short_dndm}), orbital heating scales as $\langle \Delta E \rangle \propto t^2$ on short time-scales $\omega t\lesssim 1$ independently of the size of the stellar system. However, 
the long-term evolution of $\langle \Delta E\rangle$ is more difficult to interpret. Fig.~\ref{fig:de10} shows that Equation~(\ref{eq:delv2_long_dndm}) is reasonably accurate for stellar systems deeply embedded in the DM halo, i.e. $\rh(t=0)\ll c_h$, such as Leo V. In these systems, the mean orbital energy increases steadily with time, $\langle \Delta E \rangle \propto t$, as expected from Chandrasekhar's theory.
In contrast, extended galaxy models with large initial half-light radii (e.g. Fornax and Sculptor) follow an energy evolution that plateaus as the galaxy half-light radius approaches the halo peak velocity radius, clearly departing from the linear growth expected from stochastic heating.

 The departure from the linear growth of energy, $\langle \Delta E \rangle \propto t$, likely arises from a combination of effects. 
On the one hand, setting a constant subhalo number density to $g=g(c_h)$ may not be valid at large distances $r \gg c_h$. As stars diffuse to the outer-most regions of the halo, they encounter fewer and fewer perturbers, which leads to a lower heating rate, and possibly contributes to the departure from a linear energy growth of $\langle \Delta E \rangle$ seen in Fig.~\ref{fig:de10} at late times.
On the other hand, the flattening of $\langle \Delta E \rangle$ may also point to a breakdown of the assumptions underlying the stochastic theory. (i) In Chandrasekhar's theory, the energy injection rate is proportional to the {\it local} number density of subhaloes, $g(r)$. However, this approximation fails both at large and small radii. At small radii, the number of subhaloes is limited by finite sampling and low-number statistics (see Fig.~\ref{fig:orb}). This leads to a systematic underestimate of subhaloes compared to the analytical expectation for $g(r)$. At large radii, stars feel force fluctuations that are dominated by distant subhaloes moving through the inner regions of the galaxy, which no longer act as local perturbers. (ii) Furthermore, Chandrasekhar's theory assumes that the perturbations are {\it random}, and that each star responds independently to those perturbations. Although this assumption holds reasonably well for stars deeply embedded in the potential, it worsens in the outer regions of the galaxy where stars tend to respond coherently to non-local perturbations generated by distant subhaloes. The results plotted in Fig.~\ref{fig:de10} suggest that non-locality may affect how energy is injected into stellar orbits, and call for a follow-up dedicated analysis to clarify this issue.


The plateau in $\langle \Delta E \rangle$ marks a fundamental ceiling set by the host halo structure. Galaxies cannot be heated above the peak velocity dispersion, $\sigma_{\rm max}$ (see Fig.\ref{fig:rhsig}). As a result, there is a maximum kinetic energy that stars can absorb and a corresponding lower bound to the potential energy, which occurs when the system expands to the peak circular velocity radius, $\rh = r_{\rm max}$ (see Fig.\ref{fig:W}).
To quantify this, let us define the mean energy change as $\langle \Delta E \rangle = \langle E \rangle - \langle E_0 \rangle$, where the subindex `0' refers to the initial state. For galaxies deeply embedded in the halo at $t = 0$, the initial energy is roughly $\langle E_0 \rangle \approx -GM_h / c_h$ (as in the Leo V models).
As the system expands and approaches $\rh\approx r_{\rm max}$, the mean energy approaches $\langle E\rangle_{\rm max}\approx W(\rh=r_{\rm max})=-\sigma_{\rm max}^2=0.0735\,GM_h/c_h$ (see Fig.~\ref{fig:W}). Hence, the mean variation of energy becomes $\langle\Delta E\rangle=(1-0.0735)\,GM_h/c_h \approx 0.925\,GM_h/c_h$. 
Beyond this point, the system becomes increasingly diffuse, the potential energy asymptotically vanishes, and the energy gain saturates at $\langle\Delta E\rangle\to GM_h/c_h$.
This limit, shown as blue dot-dashed lines in Fig.~\ref{fig:de10}, represents the maximum energy that stars can absorb through subhalo encounters, thus explaining the flattening seen in the long-term evolution of $\langle \Delta E \rangle$.

Finally, we note that the plateau in $\langle \Delta E\rangle $ could, in principle, arise from evaporation - that is, stars gaining sufficient energy to become unbound and thus no longer contributing to the heating process. While this mechanism is plausible, we find that evaporation is negligible in our models. Even in the stellar components that experience the strongest expansion, i.e. in the lowest-mass halo with $M_h=3\times 10^8 M_\odot$, fewer than $\sim 1\%$ of the stellar particles become gravitationally unbound.
In our models, evaporation is suppressed because energy injection becomes increasingly inefficient as stars expand beyond the peak velocity radius, as shown in the following Section.

\subsubsection{Relaxation time}\label{sec:relaxt}
The flattening of $\langle \Delta E \rangle$ with time has important implications for the long-term dynamical evolution of stars in dSphs. In particular, it signals that the relaxation process becomes progressively less efficient as the system expands. This can be quantified by estimating the relaxation time of the stellar component embedded within the dark matter halo.
Following the classical definition (e.g. Spitzer 1958), the relaxation time is defined as the time $t_{\rm rel}$ at which the variance of velocity impulses equals the velocity dispersion of the system, i.e. $\langle |\Delta{\mathbfit v}|^2\rangle(t_{\rm rel})=\sigma^2$. From Equation~(\ref{eq:delv2_long_dndm}), this yields 
\begin{align}\label{eq:trel}
    t_{\rm rel}= \sqrt{\frac{3}{32\pi^3}}\frac{\sigma^2\langle v^2\rangle^{1/2}}{B_0 \,g\, (\ln \Lambda -1.9)}\frac{3-\alpha}{G^2\msol^\alpha(M_2^{3-\alpha}-M_1^{3-\alpha})}.
\end{align}
Notice that in CDM models with $M_2\gg M_1$ and $\alpha<3$ one has that $M_2^{3-\alpha}-M_1^{3-\alpha}\approx M_2^{3-\alpha}$, which means that the low-mass end of subhalo mass function has a negligible influence on stellar heating. Instead, we find that 
the relaxation time~(\ref{eq:trel}) strongly depends on the spatial segration of stars within the DM halo at $t=0$:
\begin{itemize}
  \item For $a \ll c_h$ the stellar velocity dispersion can be approximated as $\sigma^2/\sigma_{\rm sub}^2=12\,(a/c_h) \ll 1$ (see \S\ref{sec:evol}), whereas the mean relative velocity between stars and subhaloes can be written as $\langle v^2\rangle=3\sigma_{\rm sub}^2+3\sigma^2\approx 3\sigma_{\rm sub}^2=(1/6)v_{\rm max}^2$. The number density of subhaloes is set to $g\approx g(c_h)= (16\pi c_h^3)^{-1}$ for a better comparison with the values derived from Fig.~\ref{fig:de10}. Inserting~(\ref{eq:N}) into~(\ref{eq:trel}) and setting $\rh=1.305\,a$, $v_{\rm max}^2=(1/4)GM_h/c_h$, $\alpha=1.9$ and $\xi_1\lll \xi_2$ yields 
  \begin{align}\label{eq:trelCDM_small}
  t_{\rm rel,CDM}\simeq 0.70\,\frac{\rh \,c_h^{1/2}}{(GM_h)^{1/2}\,N\,(\ln \Lambda-1.9)} \,\frac{1}{\xi_1^{0.9}\xi_2^{1.1}}~~{\rm for}~~~ a \ll c_h ,
  \end{align}

\item For $a \gtrsim c_h$ stars have reached their peak velocity dispersion, $\sigma\approx \sigma_{\rm max}=0.54\,v_{\rm max}$, thus the mean relative velocity between stars and subhaloes becomes $\langle v^2\rangle=3\sigma_{\rm sub}^2+3\sigma^2\approx 1.04\,v_{\rm max}^2$. Following similar steps as above yields
   \begin{align}\label{eq:trelCDM}
  t_{\rm rel,CDM}\simeq 0.15\,\frac{c_h^{3/2}}{(GM_h)^{1/2}\,N\,(\ln \Lambda-1.9)} \,\frac{1}{\xi_1^{0.9}\xi_2^{1.1}} ~~{\rm for}~~~ a \gtrsim c_h . 
  \end{align}
  
\end{itemize}
These equations show a few points of interest. 
(i) For galaxies deeply segregated in the halo ($\rh \ll c_h$), the relaxation time scales with the local dynamical time of stars at $\rh$, i.e. $t_{\rm rel,CDM}\sim \rh/(GM_h/c_h)^{1/2} \sim \rh /\sigma_{\rm sub}$.
In contrast, for extended galaxies ($\rh \gtrsim c_h$), relaxation is governed by the dynamical time of the host halo, $t_{\rm rel,CDM}\sim c_h^{3/2}/(G M_h)^{1/2}\sim \Omega_h^{-1}$. This implies that small galaxies with $\rh\ll c_h$ -like Leo V- are expected to have a shorter relaxation time and a faster gravothermal expansion than extended galaxy models such as Fornax or Sculptor.
Once galaxies reach the scale radius of the host, the rate at which stellar orbits evolve is linked to the large-scale properties of the host potential, losing its memory of their initial half-light radius.
(ii) In CDM simulations, the mass and scale radius of DM haloes are strongly correlated. Adopting the size-mass relation~(\ref{eq:c}) yields $\Omega_h^{-1}\sim M_h^{(3\beta-1)/2}$.  Taking $\beta \simeq 0.5$ (Erkal et al 2016), and keeping the number of subhaloes ($N$) and the limits of the subhalo mass function ($\xi_1$, $\xi_2$) fixed, we obtain
$ t_{\rm rel,CDM} \sim M_h^{0.25}$. Thus, more massive halos have longer relaxation times, suggesting that galaxies lower-mass halos expand faster than those embedded in larger haloes.
(iii) The relaxation time is inversely proportional to the number of subhaloes, $\Omega_h t_{\rm rel,CDM} \sim 1/N.$ This is characteristic of a stochastic heating process, where the cumulative effect of numerous weak perturbations dominates the orbital diffusion of stars. This behavior differs from the classic two-body relaxation time, which scales as $ \Omega_h t_{\rm rel,2b} \sim N/\ln N$.
In contrast to two-body relaxation, where increasing $N$ slows evolution, a richer subhalo population enhances the gravothermal expansion process in dSphs.

\begin{figure}
\begin{center}
\includegraphics[width=76mm]{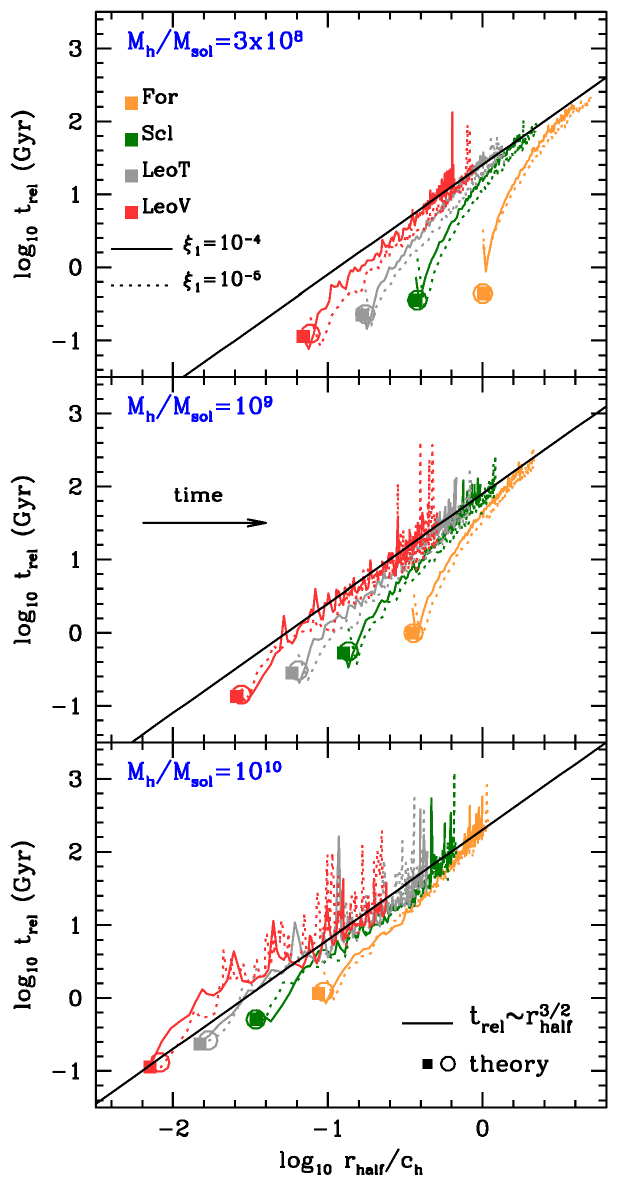}
\end{center}
\caption{Relaxation time as a function of the half-light radius, $\rh(t)$, for three different dark matter halo masses. Dotted and solid lines show experiments with low-mass truncations of the subhalo mass function at $\xi_1=10^{-5}$ and $\xi_1=10^{-4}$, respectively. The relaxation time can be fitted by a power-law scaling, $t_{\rm rel} \propto \rh^{3/2}$ (black-solid lines). In contrast, Equation~(\ref{eq:trel}) (black squares/open circles for models with $\xi_1=10^{-4}$ and $10^{-5}$, respectively) provides a reasonable estimate of the numerical values at early times, but systematically underestimates the relaxation time as the galaxies expand. This result suggests that orbital diffusion slows down as the galaxy size approaches the halo scale radius, with energy fluctuations becoming increasingly correlated at large radii. }
\label{fig:trel}
\end{figure}

The classical definition of relaxation time assumes that $t_{\rm rel}$ remains constant with time. A generalized definition that allows for the time-variation of this quantity can be written as $t_{\rm rel}=|\d \langle |\Delta{\mathbfit v}|^2\rangle/\d t/ \sigma^2|^{-1}=|\d \langle \Delta E\rangle/\d t /(\sigma^2/2)|^{-1}$.
Figure~\ref{fig:trel} shows the evolution of the (time-varying) relaxation time as a function of the half-light radius, $\rh(t)$, for the three different dark matter halo masses considered in this study. The relaxation time is measured from the heating rates plotted in Fig.~\ref{fig:de10} averaged over the subhalo ensembles.
In the static regime ($t\lesssim T_0$), we observe a slight decrease of the relaxation time as the energy impulses grow as $\langle \Delta E\rangle \propto t^2$. This is quickly followed by a rapid increase of $t_{\rm rel}$ as the heating rate starts to deviate from the stochastic behaviour $\langle \Delta E\rangle\propto t$ at $t\gtrsim T_0$.

Extending the subhalo mass function to lower masses from $\xi_1 = 10^{-4}$ down to $\xi_1 = 10^{-5}$ has little effect on the relaxation times of the dwarf galaxies. Recall that in our experiments the normalization of the subhalo mass function $B_0$ is the same across all models, which means that the total number of subhaloes increases as $N \propto \xi_1^{-0.9}$ (see Equation~\ref{eq:N}). Despite this increase, the additional low-mass subhaloes contribute very little to the heating of stellar orbits. As suggested by Equation~(\ref{eq:trel}), and confirmed by our results, the relaxation time is largely set by the most massive subhaloes, which dominate the energy transfer to stars.

The relaxation times measured from our experiments lie above the predictions from Chandrasekhar's theory of random force fluctuations, Equation~(\ref{eq:trel}) (open and closed symbols). This Equation is relatively accurate at early times of the evolution, but underestimates the numerical values of $t_{\rm rel}$ as $\rh$ expands. This is consistent with the results plotted in Fig.~\ref{fig:de10}, which show that the disagreement between the theory and the numerical results systematically worsens as the galaxy approaches its peak velocity dispersion.

Importantly, as the orbital energy saturates $\langle \Delta E\rangle \to \langle \Delta E\rangle_{\rm max}$ the relaxation time asymptotically converges towards a power-law scaling relation, $t_{\rm rel} \propto \rh^{3/2}$ (black-solid lines), independently of the halo mass or the initial spatial segregation of the stellar component.


The implications of this result are significant. Since the relaxation timescale governs the heating rate, an increasing $t_{\rm rel}$ suggests that the gravothermal expansion of the stellar component becomes self-limiting. In addition, the fact that orbital diffusion slows down with expansion limits the rate of orbital evaporation from the dwarf galaxy.


\begin{figure}
\begin{center}
\includegraphics[width=80mm]{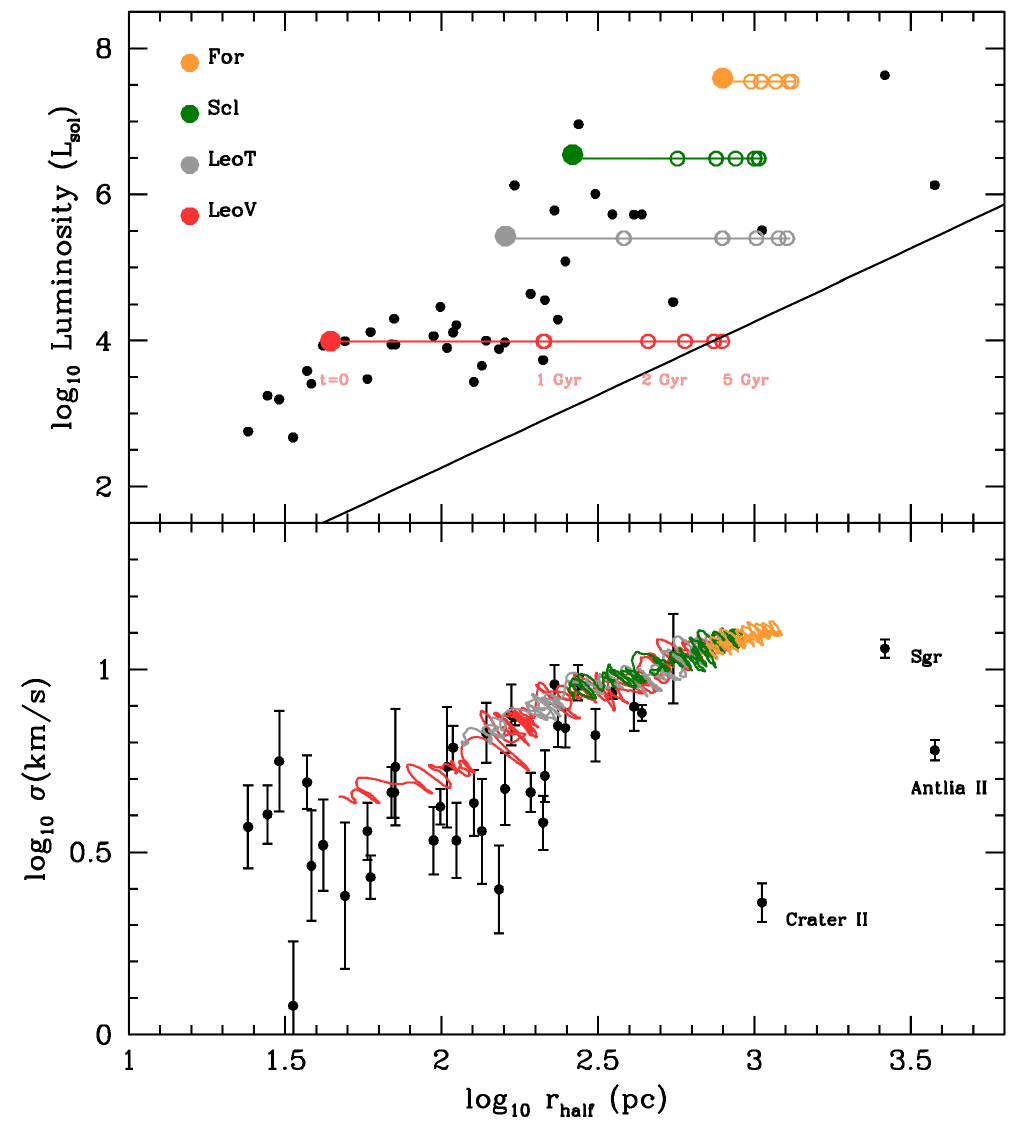}
\end{center}
\caption{{\it Upper panel:} Luminosity vs half-light radii of the Milky Way dSphs with measured velocity dispersion. Coloured lines show the evolution of the half-light radius of the models plotted in Fig.~\ref{fig:rhsig} in a fiducial Hernquist halo with a mass $M_h=10^9\msol$ and a scale radius $c_h=2.26\kpc$. For illutration, these models are integrated for 5 Gyr, with open circles marking 1 Gyr intervals. Notice that the expansion of dSphs tends to slow down as the stellar half-light radius approaches the halo scale radius. The solid-black line highlightes a surfice-brightness of 32 magitudes/arcsec$^2$, below which current photometric surveys struggle to detect dSph galaxies. The expansion of ultra-faint dSphs ($L\lesssim 10^5\,L_\odot$) can push these galaxies over the detection threshold after $\sim 2\gyr$ of evolution in a clumpy DM halo. {\it Bottom panel:} Velocity dispersion vs half-light radii of the galaxies and the $N$-body models shown in the upper panel. As expected, galaxies that expand within their DM haloes tend to become hotter as a result. After 5 Gyr of evolution, all models end up with a velocity dispersion that is close to its peak, $\sigma\simeq 0.54\,v_{\rm max}\simeq 10\kms$. Notice that during their internal expansion, dSphs roughly tracer the scaling relationship observed in the Milky Way dSPhs. As pointed out in \S\ref{sec:haloes}, Sagittarius, Antlia II and Crater II dSphs have large half-light radii and low velocity dispersions that clearly stand out from the scaling relation.}
\label{fig:rhlum_sig}
\end{figure}

\section{Discussion}\label{sec:dis}

\subsection{The evolution of dSphs towards UDGs}\label{sec:udiff}
Our results indicate that dSphs embedded in CDM haloes that host a sizeable population of dark subhaloes tend to expand monotonically over time. This finding has several important implications.

First and foremost, it suggests that the observed properties of individual dwarf galaxies, regardless of whether they are in the field or satellites of a larger galaxy, do not simply reflect the conditions imprinted by star formation at high redshift but are instead evolving with time. Stars in galaxies with old stellar populations are likely to have experienced some degree of dynamical evolution due to interactions with dark subhaloes.

Fig.~\ref{fig:rhlum_sig} shows the (future) evolution of the dSphs models embedded within a fiducial halo with $M_h=10^9\msol$ in the space of observables (see Fig.~\ref{fig:rhsig}). The upper panel plots the half-light radius ($\rh$) as a function of luminosity ($L$) for Milky Way dSphs. A solid-black line marks a constant surface brightness of 32 mag/arcsec$^2$, which approximately corresponds to the detection limit of current photometric surveys (e.g. Simon 2019). Since their stellar populations are old, dSph luminosities are assumed to remain constant, meaning that expansion drives them horizontally in this diagram. 
Over time, some dSphs cross the detection threshold, becoming too faint to be observed. However, as $\rh$ grows the heating mechanism that drives the expansion becomes less efficient (see Fig.~\ref{fig:trel}), which leads to a progressively size stalling. To illustrate this, we mark intervals of $1\gyr$ over a $5\gyr$ period, showing how the expansion rate decreases with time. This effect primarily affects the ultra-faint galaxies ($30\lesssim \rh/\pc\lesssim 100$). Notably, the model for Leo V crosses the detection threshold in just $\sim 2\gyr$ of evolution, which is a small fraction of the stellar ages in these galaxies (age$\gtrsim 12\gyr$, e.g. Grebel \& Gallager 2004)\footnote{Recall that the models plotted in Fig.~\ref{fig:rhlum_sig} assume that dSphs are embedded in a fiducial dark matter halo with mass $M_h=10^9\msol$ (see \S\ref{sec:haloes}). If dSphs reside in haloes of different masses, the associated timescales can vary significantly across the population.}

The lower panel shows the evolution of $\rh$ versus velocity dispersion, $\sigma$. As dSphs expand, they also heat kinematically, following a trajectory roughly parallel to the observed $\rh$-$\sigma$ relation\footnote{
Note that subhalo-driven expansion cannot explain the extreme size and low velocity dispersion of dSphs like Antlia II and Crater II, which remain open puzzles (Torrealba et al. 2016, 2019). In addition, the Sagittarius (Sgr) dwarf galaxy is currently undergoing tidal disruption and in disequilibrium (Ibata et al. 1994)
}. This complicates a distinction between galaxies that have expanded over time and those that formed with their current size. 

Our analysis suggests the presence of a population of faint, extended satellites that remain undetectable in current photometric surveys - they would be galaxies with large half-light radii, low luminosities, low metallicities and relatively high velocity dispersions that scale directly to the peak velocity of the host halo, $\sigma\approx \sigma_{\rm max}\simeq 0.54,v_{\rm max}$.
These objects bear some similarities to the `stealth' galaxies predicted by Bullock et al. (2010), with the difference that while stealth galaxies form with large sizes, our dSph models initially have small sizes, which gradually expand over time.
Dwarf galaxies that cross the detection threshold in Fig.~\ref{fig:rhlum_sig} would have sizes comparable to Ultra-Diffuse Galaxies (UDGs), which are low-surface brightness systems ($> 24$ mag/arcsec$^2$) with large effective radii ($\rh >1.5 \kpc$), and relatively high velocity dispersions (e.g. van Dokkum et al. 2016). However, a key distinction is that UDGs are significantly more luminous ($L \gtrsim 10^8 L_\odot$), whereas the ultra-faint dSphs plotted in this figure are several orders of magnitude fainter, $L\lesssim 10^5 L_\odot$.

It is also worth noting as dSphs expand within their DM haloes, they become less resilient to stellar tidal stripping (e.g. Pe\~narrubia et al. 2008b). Once they start to lose stars to tides, the stellar luminosity, velocity dispersion, central surface brightness, and half-light radius would tend to decrease with time (e.g. Errani et al. 2022). This effect complicates the interpretation of the scaling relationships shown in Fig.~\ref{fig:rhlum_sig}, as the observed trends may reflect not only internal secular evolution but also varying degrees of tidal disruption.

\subsection{Disequilibrium features}\label{sec:diseq}
Although the expansion of dSphs proceeds in a quasi-virial state, transient perturbations typically arise during the pericentric passages of the most massive subhaloes in the galaxy. These encounters inject energy into the system in a highly non-uniform manner, producing short-lived disequilibrium features that disrupt the otherwise smooth, self-similar evolution of the galaxy. As illustrated in Fig.~\ref{fig:xyz}, such perturbations generate shell-like structures and overdensities, where stars displaced from equilibrium temporarily accumulate in coherent structures before phase mixing. The stellar distribution also develops global asymmetries, as the fluctuating force field distorts and flattens the galaxy shape, stretching and compressing different regions in a time-dependent manner. These morphological disturbances are accompanied by kinematic signatures, including localized increases in velocity dispersion and transient bulk flows that deviate from equilibrium expectations. In some cases, stars gain enough energy to populate extended, loosely bound envelopes, further complicating the identification of accreted stellar haloes (e.g. Yang et al. 2022).

The kinematic response of stars to these perturbations may provide an avenue for testing the presence of dark subhaloes, as velocity gradients, and spatially correlated velocity substructures could be observable tell-tales of subhalo interactions. In a follow-up study, we will systematically investigate these kinematic imprints and explore their implications for current and future spectroscopic surveys of dSphs.

\subsection{dSphs with multiple chemodynamical components}\label{sec:comp}
The discovery of dSphs with multiple chemodynamical stellar populations (e.g. Tostoy et al. 2004; Battaglia et al. 2008; Walker \& Pe\~narrubia 2011; Fabrizio et al. 2016; Kordopatis et al. 2016; Pace et al. 2020; Arroyo-Polonio et al. 2024) opens the possibility to time the expansion of these galaxies using stars with different ages. In these systems, metal-poor stars are typically older and more spatially extended and kinematically hotter than their metal-rich (younger) counterparts. This trend aligns with our findings, which show that stars in a clumpy, DM-dominated potential undergo gradual expansion and heating due to interactions with dark subhaloes. In this framework, older stellar populations have experienced prolonged dynamical heating compared to younger ones.

This may allow us to use dSphs with multiple chemodynamical populations to constrain the properties of dark subhaloes in those galaxies by modelling the positions, velocities, and ages of each stellar component separately. However, a key limitation is the uncertainty in the properties of the stellar populations at the time of formation. Specifically, the present-day size of a given population could be influenced by both the subhalo population and the initial configuration of stars in the DM halo. This degeneracy between the initial structure of the galaxy and the subhalo mass function presents a challenge that requires further analysis.
For instance, if the extended, metal-poor component of dwarf galaxies such as Fornax or Sculptor has been significantly heated by subhaloes, we might expect: (i) signs of non-equilibrium dynamics (as discussed in Section~\ref{sec:diseq}), (ii) a reduced number of wide binaries, as these are fragile systems susceptible to disruption by subhaloes (Pe\~narrubia et al. 2010b), and (iii) a velocity dispersion that becomes radially anisotropic at large radii (see Fig.~\ref{fig:distrib}). These are some examples  of how future data may help to break the formation vs evolution degeneracy and test the theoretical scenario proposed in this work.

\subsection{The puzzling (small) sizes of ultra-faint dSphs}\label{sec:ufaint}
Our models suggest that if dSphs currently host bound dark subhaloes, then these galaxies must have been significantly more compact in the past. This effect is particularly pronounced for the ultra-faint dSphs, which undergo the most rapid expansion in Fig.~\ref{fig:rhlum_sig}. Notably, the ultra-faint dSph model becomes effectively `undetectable' within $\sim 2\gyr$ of evolution in a clumpy halo—an outcome at odds with their observed ancient stellar populations ($\gtrsim 12 \gyr$; e.g. Simon 2019). This raises the possibility that the small physical sizes of ultra-faint dSphs are in tension with the CDM framework. However, while our statistical experiments provide valuable physical insight, they remain highly idealized and do not capture the full complexity of dSph evolution in a realistic Galactic environment. Several key limitations should be noted:

\begin{itemize}
 \item   {\it Static host halo:} Our models assume a fixed dark matter potential. Yet, dSphs in a cosmological setting undergo an initial phase of hierarchical growth -where their mass and size increase- followed by tidal stripping after accretion onto the Milky Way, which leads to mass loss and size contraction (e.g., Pe\~narrubia et al. 2010a). Additionally, tidal stripping and baryonic feedback may modify the density profile of these galaxies, e.g. by removing the primordial DM cusp (e.g., Navarro et al. 1996; Pontzen \& Governato 2012; Pe\~narrubia et al. 2012; di Cintio et al. 2014). Modelling these effects self-consistently requires dedicated simulations of dSphs in a cosmological framework.

\item {\it Static subhaloes:} Our subhaloes also source static potentials. This is motivated by Errani \& Navarro (2021), who show that subhaloes in a tidal field evolve toward an asymptotic density profile that remains time-invariant. However, by construction our models assume that subhaloes have already reached the asymptotic limit at $t=0$, thus not capturing the dynamical process by which these subhaloes reach this asymptotic state via mass loss an subsequent re-virialization. Following the dynamical evolution of subhaloes within satellite galaxies is currently an open problem in computational cosmology, as discussed in the Introduction.

\item {\it Subhalo survival:} Whether subhaloes can be fully disrupted by galactic tides remains an open question. The asymptotic evolution found by Errani \& Navarro (2021) apply to subhaloes generated with isotropic velocities, but Chiang et al. (2024) recently showed that subhaloes with radially anisotropic velocity distributions suffer enhanced tidal stripping and may experience a cusp-to-core transformation that leads to complete disruption. 

\item {\it Major mergers:} Our models neglect the role of large subhalo mergers ($M/M_h\gtrsim 10^{-1}$), which, due to dynamical friction, sink toward the centre of the dwarf galaxy. While these events are rare and occur primarily during the early hierarchical growth phase (before dSph accretion onto the Milky Way), they can introduce significant perturbations to the galaxy. Also, if these objects experience in-situ star formation, they can contribute to the formation of a stellar halo in dSphs (e.g., Revaz 2023).


\item {\it Heating of the dark matter host halo:} Our models focus on the dynamical response of stellar tracers, but the host dark matter halo may also be subject to heating by dark subhaloes. As shown by El-Zant et al. (2016), stochastic fluctuations can transfer energy to a cuspy host halo, potentially leading to the formation of a DM core. On the other hand, DM cusps can also re-grow via subhalo mergers (e.g. Laporte \& Pe\~narrubia 2015). The key distinction between the dark matter and stellar components lies in their formation histories. Most stars in dSphs form early and over a short timescale, making them susceptible to heating over cosmic time. In contrast, dark matter can be continuously deposited in the central regions via mergers, enabling cusp regeneration. In this framework, the present-day dark matter profile is the outcome of a hierarchical mass assembly rather than the final stage of secular dynamical evolution.

\item {\it Unknown dark matter nature:} We assume dark matter is collisionless. However, if DM is made of Primordial Black Holes (PBHs) or self-interacting particles (SIDM) subhaloes can undergo a cusp-core transformation, followed by a rapid gravothermal collapse, which can lead to extremely cuspy inner profiles or, in extreme cases, even the formation of a central black hole. If the self-interaction cross section is velocity-dependent, the present-day population of subhaloes as well as the host haloes may exhibit a wide diversity of density profiles (e.g. Col\'in et al. 2002; Meshveliani et al. 2022).
  
\item {\it Subhalo abundance:} We assume that dSphs retain the same relative abundance of subhaloes as field haloes. While this assumption allows us to re-scale the subhalo mass function found in the Aquarius simulations down to the mass-scale of dSphs, it neglects the effects of Galactic tides, which may strip a fraction of the subhaloes bound to these galaxies at their time of accretion onto the MW. Currently, cosmological simulations provide no clear predictions for the number, distribution, and mass function of subhaloes orbiting in the MW dSphs. This is complicated by resolution limits, artificial disruption, and inconsistencies in subhalo identification within the parent haloes. In this sense, semi-analytical codes such as SatGen (Jiang et al. 2021) are valuable alternative tools to study this issue in more detail.

 \item {\it Subhalo spatial distribution:} In this work, we assume that subhaloes trace the underlying density profile of the smooth host halo. The validity of this approximation is unclear. Early simulations of structure formation showed that subhaloes follow a cored number density profile (see e.g., Diemand, Moore \& Stadel 2004; Springel et al. 2008). In contrast, using semi-analytical models, Green, van den Bosch \& Jiang (2021) found that, at least to some extent, the deficit of subhaloes at small radii may a numerical artifact arising from both limiting mass resolution and artificial disruption (see also Santos-Santos et al. 2024). The unclear spatial profile of the subhalo population adds uncertainty when it comes to estimating the expansion rate of dSphs in a cosmological context. E.g. a cored number density profile $g(r)$ may slow down gravothermal expansion, and thus help to explain the existence of DM-dominated ultra-faint dSphs with $r_{\rm half}\sim 30$pc and stellar ages $\gtrsim 12$Gyr.  
  
\item {\it No stellar self-gravity:} In our models, stars are treated as test particles in the dark matter potential. However, if dSphs formed as compact stellar clusters within a host DM halo, stellar self-gravity may also play an important role in slowing down gravothermal expansion (Brandt 2016). 
 In such cases, self-gravity introduces a characteristic {\it dissolution timescale}, $t_{\rm dis}$, which measures the lifetime of a self-gravitating cluster in a clumpy medium. We expect subhaloes to heat stellar clusters, increasing both their size and internal velocity dispersion, until they become gravitationally unbound. Once dissolution occurs, stellar self-gravity becomes negligible, and stars expand in the host halo under the influence of subhalo-induced force fluctuations. Brandt (2016) studied  how the lone cluster in Eridanus II dSph is heated by MACHOs, finding that the dissolution timescale lies in the range $t_{\rm dis}\sim 1$--$6\gyr$ depending on the MACHO density.

\end{itemize}
  These limitations highlight several challenges in reconstructing the past evolution of dSphs, particularly in understanding the mechanisms that led to the observed compactness of ultra-faint systems. Addressing these issues goes beyond the scope of this paper and will require additional theoretical work to incorporate more realistic cosmological environments and an improved dynamical modelling of the stellar \& subhalo populations in dSphs.
  
Despite the idealized nature of the experiments presented in this paper, the results shown here are broadly consistent with the evolution of dwarf galaxies found in cosmological hydrodynamical simulations. While these simulations successfully reproduce the global properties—luminosity, velocity dispersion, and star formation histories—of brighter Milky Way dSphs ($L \gtrsim 10^5\,L_\odot$), they consistently struggle to reproduce the compact sizes observed in the ultra-faint population ($L \lesssim 10^5\,L_\odot$). 
 This is because ultra-faint galaxies ($L \lesssim 10^5, L_\odot$) tend to experience a significant amount of expansion after forming the bulk of their stars. 
As a result, the half-light radii for simulated ultra-faint galaxies in cosmological hydrodynamical simulations are typically an order of magnitude larger than those observed (Revaz \& Jablonka 2018; Wheeler et al. 2019; Agertz et al. 2020; Applebaum et al. 2021; Orkney et al. 2021, 2023; Gutcke et al. 2022).

Interestingly, a recent study by Revaz (2023) combining cosmological hydrodynamical simulations with high-resolution dark-matter-only runs showed that ultra-faint galaxy expansion is primarily driven by accretion of substructures. This led to the conclusion that {\it the difficulty in reproducing the compactness of ultra-faint dSphs arises from the hierarchical nature of small-scale structure formation in CDM,} noting that this tension could potentially be alleviated in Warm Dark Matter (WDM) models that suppress the number of subhaloes below a mass scale dictated by the DM particle mass.

  \subsection{Other sources of heating}
Yet, there are several sources of gravitational fluctuations in dSphs that may induce comparable effects. For example, in CDM models feedback-driven gas motions or orbiting baryonic clumps can transfer energy to DM particles and drive cusp-core transformations (e.g. Pontzen \& Governato 2012; Muni et al. 2025). These fluctuations, which can be described by a power-law power spectrum characteristic of turbulent media, act as a diffusive heat source, which in principle can be modelled as a generalized form of Chandrasekhar's theory of two-body relaxation, where the coupling strength is set by the gas fraction and fluctuation amplitude (e.g. El-Zant et al. 2016; Hashim et al. 2023). Similarly, in fuzzy dark matter (FDM) scenarios, wave interference patterns naturally generate time-dependent density fluctuations that heat embedded stellar systems over gigayear timescales, leading to expansion, radial anisotropy, and morphological evolution (e.g. El-Zant et al. 2020; Dutta Chowdhury et al. 2023). 

Our results suggest that, irrespective of the specific heating mechanism, the long-term evolution of dSphs is governed by the same gravothermal principles. In particular, energy injection - whether from dark subhalo encounters, baryonic feedback, or quantum interference - leads to a net expansion of the stellar component within the DM halo. As a result, the velocity dispersion increases with time at all radii but more so closer to the center, and the velocity anisotropy becomes radially biased at large radii.
As the dSph half-light radius approaches the peak velocity radius of the host dark matter halo, further increases in stellar velocity dispersion are suppressed. This point marks a thermodynamic phase transition, where the sign of the heat capacity changes from positive to negative. Once the heat capacity becomes negative, additional energy input no longer leads to a significant temperature increase but instead drives expansion. 
This saturation effect is expected to be generic to stellar tracers orbiting finite-mass haloes and imposes a limit on the internal velocity dispersion of dSphs. While our analysis focuses on heating by subhalo encounters, the appearance of this phase transition may be a generic feature of tracer populations acted on by random force fluctuations in a DM potential.

\section{Summary }\label{sec:sum}

In this work, we have investigated the dynamical response of dwarf spheroidal galaxies (dSphs) embedded in DM haloes that host a population of subhaloes. Our key findings can be summarized as follows:

\begin{itemize}
\item Dark subhaloes generate a combined fluctuating force that injects energy onto stellar orbits, which causes a \textit{gradual expansion} of dSphs. However, this expansion is neither smooth nor continuous. Instead, it is driven by massive subhaloes passing through pericentre, which dominate the energy injection. The effects of these interactions are transient, often leading to temporary elongations or shell-like structures that vanish once the system settles into a new equilibrium state characterized by a larger half-light radius.

\item  Despite the overall expansion, the stellar density profile, $\nu(r)$, remains close to its initial shape, suggesting a self-similar evolution. 
  
\item In contrast, the velocity dispersion profile, initially flat, $\sigma(r) \sim \text{const}$, evolves as the galaxy expands. The inner regions become \textit{kinematically hotter}, while the outer regions \textit{cool down}.

\item The expansion rate declines with time as $\rh$ approaches the peak velocity radius of the halo ($\rh \approx r_{\rm max}$). The evolution of the luminosity-averaged velocity dispersion is more complex: at early times, $\sigma$ increases as the galaxy expands, but once $\rh$ reaches $r_{\rm max}$, the velocity dispersion peaks at the virial limit $\sigma_{\rm max} \simeq 0.54 v_{\rm max}$. As $\rh$ exceeds the peak velocity radius at late times, $\sigma$ begins to decline.
    
\item The heat capacity of the stellar system remains positive for stars deeply segregated within the DM potential, but \textit{diverges} as the half-light radius approaches the peak velocity radius. Beyond this threshold, the heat capacity becomes negative, indicating a transition in the dynamical response to energy injection. This is reminiscent of second-order phase transitions in thermodynamics, where a critical point separates two distinct regimes of energy redistribution.

\item The dynamical evolution observed in these models shares key similarities with gravothermal collapse in self-gravitating systems (e.g. Lynden-Bell \& Wood 1968), but follows an \textit{inverted} trajectory. As the stellar system expands within a clumpy dark matter potential, the inner regions heat up (i.e. their velocity dispersion increases), while the outer regions cool down, leading to a steepening of the velocity dispersion profile over time. This redistribution of kinetic energy mirrors the kinematic signature of gravothermal collapse, but takes place in galaxies expanding in a DM potential. The emergence of a heat capacity divergence and correlated energy fluctuations - particularly as the galaxy size approaches the peak velocity radius of the DM halo - suggests that the underlying dynamical mechanism deviates from a purely stochastic heating process.

\item  The relaxation time increases as $t_{\rm rel} \propto \rh^{3/2}$, independently of halo mass or the initial segregation of the stellar component. 
This deviates from a constant relaxation time expected from stochastic heating, and suggests that orbital diffusion slows down as the galaxy expands. Physically, this slow -down happens because at large radii both stars and subhaloes move very slowly and the density of subhaloes is low, which decreases the probability of close encounters. Instead, force fluctuations are generated by distant subhaloes moving through the inner -most regions of the halo. This behaviour cannot be well reproduced by Chandrasekhar's local stochastic theory, which assumes that encounters are dominated by nearby objects moving on random straight -line trajectories in a uniform background.

  \item Observationally, our results suggest that the present-day properties of dSphs are not merely relics of early star formation but reflect ongoing secular evolution. Extrapolating our models forward in time, we find that some ultra-faint dSphs may expand beyond the detection limit of current photometric surveys, thus becoming `stealth' galaxies. These systems have large sizes $\rh \sim r_{\rm max}$, relatively high velocity dispersions, $\sigma\sim \sigma_{\rm max}$, and low surface brightness, thus sharing some characteristics with Ultra-Diffuse Galaxies (UDGs), though with significantly lower luminosities and metallicities.

\item The presence of multiple chemodynamical stellar populations in dSphs provides an opportunity to study their expansion history. In these galaxies, older, metal-poor stars tend to be more spatially extended and kinematically hotter than their younger, metal-rich counterparts -consistent with the effects of dynamical heating. This opens the possibility of using multiple stellar populations in dSphs to constrain the subhalo mass function by separately modelling the positions, velocities, and ages of each component. However, a major challenge is the degeneracy between the initial sizes of each stellar sub-component and the influence of dark subhaloes, which may complicate direct constraints on the masses and densities of these objects.

\end{itemize}

    \section*{Acknowledgements}
The authors would like to thank Mark Gieles and Frank van den Bosch for their insightful comments and suggestions.

Eduardo Vitral acknowledges funding from the Royal Society, under the Newton International Fellowship programme (NIF\textbackslash R1\textbackslash 241973). RE and MW acknowledge support from the National Science Foundation (NSF) grant AST-2206046. Support for program JWST-AR-02352.001-A was provided by NASA through a grant from the Space Telescope Science Institute, which is operated by the Association of Universities for Research in Astronomy, Inc., under NASA contract NAS 5-03127. This material is based upon work supported by the National Aeronautics and Space Administration under Grant/Agreement No. 80NSSC24K0084 as part of the Roman Large Wide Field Science program funded through ROSES call NNH22ZDA001N-ROMAN.

\section*{Data availability}
No data were generated for this study.

{}

\end{document}